\documentclass[12pt,preprint]{aastex}
\newcommand{\unit}[1]{\ensuremath{\, \mathrm{#1}}}
\begin{document}

\title{A Significant Problem With Using the Amati Relation for Cosmological Purposes}
\author{Andrew C. Collazzi, Bradley E. Schaefer \affil{Dept.  of Physics and Astronomy, Louisiana State University, Baton Rouge, LA 70803} Adam Goldstein, Robert D. Preece  \affil{Dept.  of Physics, University of Alabama, Huntsville, AL 35899}}

\begin{abstract}
We consider the distribution of many samples of Gamma-Ray Bursts (GRBs) when plotted in a diagram with their bolometric fluence ($S_{bolo}$) versus the observed photon energy of peak spectral flux ($E_{peak,obs}$).  In this diagram, all bursts that obey the Amati relation (a luminosity relation where the total burst energy has a power law relation to $E_{peak,obs}$) must lie above some limiting line, although observational scatter is expected to be substantial.  We confirm that early bursts with spectroscopic redshifts are consistent with this Amati limit.  But, we find that the bursts from BATSE, {\it Swift}, Suzaku, and Konus are all greatly in violation of the Amati limit, and this is true whether or not the bursts have measured spectroscopic redshifts.  That is, the Amati relation has definitely failed.  In the $S_{bolo}-E_{peak,obs}$ diagram, we find that every satellite has a greatly different distribution. This requires that selection effects are dominating these distributions, which we quantitatively identify.  For detector selections, the trigger threshold and the threshold for the burst to get a measured $E_{peak,obs}$ combine to make a diagonal cutoff with the position of this cutoff varying greatly detector to detector.  For selection effects due to the intrinsic properties of the burst population, the distribution of $E_{peak,obs}$ makes for bursts with low and high values to be rare, while the fluence distribution makes for bright bursts being relatively uncommon.  For a detector with a high threshold, the combination of these selection effects serves to allow only bursts within a region along the Amati limit line to be measured, and these bursts will then appear to follow an Amati relation.  Therefore, the Amati relation is an artifact of selection effects within the burst population and the detector.  As such, the Amati relation should not be used for cosmological tasks.  This failure of the Amati relation is in no way prejudicial against the other luminosity relations.
\end{abstract}

\keywords{gamma-ray burst: general, gamma-ray: stars}

\section{Background}

Luminosity relations are a tool for which we can connect a measurable quantity (called a luminosity indicator) from a long-duration Gamma-Ray Burst (GRB) to the burst's peak luminosity or total energy.  The inverse square law and the observed burst brightness can then be used to determine the distance to the burst.  In doing this, GRBs become a `standard-candle' (in the same sense as Cepheid variables and Type Ia supernovae) where observed properties can be used to determine luminosity and then distance.  GRBs are seen out to very high distances, at least to redshift z=8.2 (Tanvir et al. 2009), and so the GRB luminosity relations offer a unique and powerful opportunity for new cosmology.

Currently, GRBs have seven well-established luminosity relations (Fenimore \& Ramirez-Ruiz 2000; Norris et al. 2000; Amati et al.  2002; Schaefer 2002; Schaefer 2003; Ghirlanda et al.  2004).  The luminosity indicators are the spectral lag time ($\tau_{lag}$, Norris et al. 2000), `variability' (V, Fenimore \& Ramirez-Ruiz 2000), the peak of the $\nu F_\nu$ power spectrum ($E_{peak}$; Schaefer 2003), the minimum rise time in the light curve ($\tau_{RT}$; Schaefer 2002), and the number of peaks in the light curve ($N_{peak}$, Schaefer 2002).  Five of the luminosity relations connect the luminosity indicators to the burst's peak luminosity.  Two luminosity relations instead connect to the total energy of the burst, with the physics relating to the burst energetics instead of the conditions at the time of peak luminosity.  These are: the (A) Amati relation (Amati et al.  2002;2006), which connects $E_{peak}$ to the total energy of the burst assuming for isotropic emission ($E_{\gamma,iso}$), and (B) the Ghirlanda relation (Ghirlanda et al.  2004) which connects $E_{peak}$ to the total energy of the burst, but with a correction for the beaming factor ($F_{beam}$), resulting in the energy emission ($E_{\gamma}$).  In addition, there are a variety of newer luminosity relations have been proposed, but not yet fully tested and confirmed by the community (e.g.  Dainotti et al.  2008; 2010; 2011) and while others have proven to not be viable.  An example of this is the Firmani relation (Firmani et al.  2006), which was found to be no significant improvement over existing relations (Collazzi \& Schaefer 2008).  Physical explanations for a variety of the established relations have been made (e.g.  Kobayashi et al.  2002; M\'{e}sz\'{a}ros et al.  2002; Schaefer 2003; Eichler \& Levinson 2004; Liang et al.  2004; Schaefer 2004;  Rees \& M\'{e}sz\'{a}ros 2005; Giannios \& Spruit 2007; Thompson et al.  2007).  

Recently, Collazzi et al.  (2011) performed an exhaustive study on the sources of error on $E_{peak}$.  This study showed a considerable amount of scatter was hidden in how $E_{peak}$ is measured, which was much larger than that from the reported Poisson errors alone.  The sources of this scatter included: (A) the choices of different analysts, (B) which $E_{peak}$ is measured (e.g. the time-integrated or time-resolve $E_{peak}$), and (C) the detector response matrix, all in addition to the regular Poisson statistical error.  This scatter can be as large as 0.43 in log space and has a typical value of 0.24.  This scatter can explain the scatter seen in the luminosity relations that use $E_{peak}$.  

Still, even the currently accepted luminosity relations have their drawbacks.  The best (i.e.  the tightest) of these relations, the Ghirlanda relation, can only be applied if there is an observed jet break.  Jet breaks are a well understood phenomena (Rhoads 1997; Sari et al. 1999).  Measuring a jet break is fairly difficult for a variety of reasons, and has only been observed in a small percentage of bursts.  Melandri et al.  (2008) and Kocevski \& Butler (2008) have pointed out problems in identifying these jet breaks with the X-ray data. 

Most notably, the Amati relation has been criticized for several reasons.  Li (2006) originated a test that demonstrated an ambiguity when the measured properties are used to determine the redshift.  This result was confirmed by other groups later (e.g.  Schaefer \& Collazzi 2007).  This ambiguity also exists for the Ghirlanda relation, but at a substantially higher redshift.  This ambiguity does not exist for any of the other confirmed luminosity relations, and so the ambiguity problem can go away if multiple relations are used to determine redshift.  This problem does \textit{not} affect current work on the GRB Hubble Diagram because those luminosity indicators were derived from a known redshift obtained spectroscopically.

Another major criticism towards the Amati relation came from Nakar and Piran (2005).  In this work, Nakar and Piran developed a test specifically for the Amati relation, the beauty of the test being that a redshift was not needed.  This test has since been generalized in several independent investigations (e.g.  Band \& Preece 2005; Schaefer \& Collazzi 2007; Goldstein et al. 2010).  They combined the Amati relation (equation \ref{eq:AmatiRel}) with the inverse square law for fluences (equation \ref{eq:ISLF}) to eliminate $E_{\gamma,iso}$ (equation \ref{eq:NaP}).
\begin{equation}
E_{\gamma,iso} = (9.2 \times 10^{47}\unit{erg \; keV^{-2.04}}) \left[E_{p}\left(1+\textit{z}\right)\right]^{2.04}
\label{eq:AmatiRel}
\end{equation}
\begin{equation}
E_{\gamma,iso} = \frac{4 \pi d_{L}^2 S_{bolo}}{1+\textit{z}}
\label{eq:ISLF}
\end{equation}
\begin{equation}
\frac{E^{2.04}_{peak}}{S_{bolo}} = \frac{4 \pi d_{L}^2}{(1+\textit{z})^{3.04} (9.2\times10^{47} \unit{erg\; keV^{-2.04}})}
\label{eq:NaP}
\end{equation}
Here, $E_{\gamma,iso}$ is the isotropic gamma ray energy, $d_L$ is the luminosity distance as derived with the concordance cosmology ($\Omega_\Lambda = 0.7$, $\Omega_m = 0.3$, H$_0$ = 74 km/s/Mpc), $S_{bolo}$ is the bolometric fluence (the fluence over the burst rest frame 1-10,000 keV range), and \textit{z} is the redshift of the burst.  The quantity $\frac{E^{2.04}_{peak}}{S_{bolo}}$ has been called the `energy ratio' for the Amati relation (e.g. Band \& Preece 2005).  The left side of equation 3 uses only directly observable quantities (albeit, they are model dependent), while the right side is only a function of distance.  As the distance rises, $d_L^2$ gets larger and $(1+\textit{z})^{-3.04}$ gets smaller, which gives a maximum value for the right side.  When the concordance cosmology is used, the function peaks at $\textit{z}\sim3.6$.  Specifically, the right side of the equation cannot exceed $1.13 \times 10^9 \unit{keV^{2.04} \; erg^{-1} \; cm^{2}}$ and, therefore, 
\begin{equation}
\frac{E^{2.04}_{peak,obs}}{S_{bolo} }  \leq 1.13 \times 10^{9} \unit{keV^{2} \; erg^{-1} \; cm^{2}}.
\label{eq:AmatiLimit}
\end{equation}
This becomes a simple way to test the Amati relation even for bursts without redshifts.  

Similarly, for the Ghirlanda relation,
\begin{equation}
E_{\gamma} = (1.35 \times 10^{47}\unit{erg \; keV^{-1.43}}) \left[E_{p}\left(1+\textit{z}\right)\right]^{1.43}
\label{eq:GhirRel}
\end{equation}
\begin{equation}
E_{\gamma} = \frac{4 \pi d_{L}^2 S_{bolo}F_{beam}}{1+\textit{z}}
\label{eq:GISLF}
\end{equation}
\begin{equation}
\frac{E^{1.43}_{peak}}{S_{bolo}} = \frac{4 \pi d_{L}^2 F_{beam}}{(1+\textit{z})^{2.43} (1.35\times10^{47} \unit{erg\; keV^{-1.43}})}
\label{eq:NaPG}
\end{equation}
The beaming factor, $F_{beam}$, is defined as $(1-\cos \theta_{jet})$, where $\theta_{jet}$ is the opening angle of the jet of the burst.   The right hand side has a maximum value at $\textit{z}_{max} = 12.6$ with a value of $2.7 \times 10^{10} \unit{keV}^{1.43} \unit{erg}^{-1} \unit{cm}^{2}$ for $F_{beam}=1$.  Thus, the Ghirlanda relation forces the limit, 
\begin{equation}
\frac{E^{1.43}_{peak,obs}}{S_{bolo} }  \leq 2.7 \times 10^{10} \unit{keV^{1.43}  \; erg^{-1} \; cm^{2}}.
\label{eq:GhirLimit}
\end{equation}
So we have a simple observational test for compliance with the Ghirlanda relation. We also have reproduced the result that the `energy ratio' for the Ghirlanda differs from the Amati relation (e.g. Band \& Preece 2005).

At first glance, it might appear that this it is being overly generous to apply a beaming factor of $F_{beam}=1$ to calculate the limit of the Ghirlanda relation.  The whole point of applying such a beaming factor is to give an illustration of the lowest value the limit can have.  If we were apply some sort of average beaming factor, the value of the Ghirlanda limit would increase, resulting in many more rejected bursts.  As an example, let us use a typical jet angle, $\theta_{jet}=8.5^{o}$, which corresponds to a beaming factor,  $F_{beam}=0.01$.  In this case, the Ghirlanda limit would increase by a power of two to $\approx 2.7 \times 10^{8} \unit{keV^{1.43}  \; erg^{-1} \; cm^{2}}$.  There are two reasons, however, for choosing to keep the Ghirlanda limit with a $F_{beam}=1.0$.  The first is the simple mathematical statement that we are looking for the maximum value for $\frac{E^{1.43}_{peak}}{S_{bolo}}$.  The second, and more physical explanation, is that the Ghirlanda relation accounts for all beaming factors, and thus the limit \textit{should} given as when $F_{beam}=1.0$.  Indeed, as we will see, the Amati relation can be thought of as a sort of Ghirlanda relation that was constructed with some average beaming factor of bursts.

Nakar \& Piran (2005) analyzed 751 \textit{BATSE} bursts, finding 48\% of the bursts to violate the Amati limit (equation \ref{eq:AmatiLimit}).  They used this result to declare the Amati relation as not reliable.  Schaefer \& Collazzi (2007) later showed that this fraction was what is expected for this relation due to simple scattering effects.  In the ideal case with no measurement uncertainties, there are no violators, but as soon as noise is introduced, some of the points (particularly those near the limit) would become violators.  Of course, there would be an equal number of bursts that would go up (and thus above the limit) to those who go down (and below the limit).  With the function on the right side of equation 3 being near its maximum value (when compared to the uncertainties in E$_{peak}$ and the scatter in the luminosity relations) for the redshift of most bursts (1 $<$ \textit{z} $<$ 6), this means that about half of the GRBs will be apparent violators even if the Amati relation is correct.  So the real finding is that the differences between the limit and the equation are very small given the known scatter (Schaefer \& Collazzi 2007).  Thus, the original test by Nakar and Piran actually \textit{confirmed} the Amati relation.

Schaefer \& Collazzi (2007) extended this test to 69 bursts with known redshifts from many satellites.  The result was that the Amati relation had an expected amount of violators, 44\%.  The paper goes on to show that this test could be generalized for use to test \textit{all} of the luminosity relations.  In most cases, this resulted in no maximum, or at the very least no maximum out to very high redshift ($\sim 20$).  All luminosity relations were shown to have either no bursts failing the test (like the lag-luminosity relation), or a number of failures that is acceptable given measurement errors).  The end conclusion is that all accepted luminosity relations of the time had passed the Nakar and Piran test.

A complication arises in Band \& Preece (2005), which examined (largely) the same BATSE bursts as Nakar and Piran without known redshifts, finding a $>$80\% failure rate for the Amati relation.  This result was later confirmed by Goldstein et al. (2010).  In particular, the spread of points was greatly over the limit, and this cannot be attributable to measurement error or scatter in the luminosity relation.  With the differences in the violator fraction and the interpretation, we have a core dilemma for the Amati relation, and the understanding of these differences is the core of this paper.  

Another criticism of the Amati relation is that the Amati relation is both dependent on the satellite and that it arises from selection effects (Butler et al.  2007).  Butler et al.  (2007) pointed out an apparent shift in the Amati relation between \textit{Swift} and pre-\textit{Swift} data sets.  Their claim that selection effects will produce the Amati relation were never substantiated by any analysis, examples, or derivations, and the cause of the selection effects was never identified.  These claims have scared off some workers from using any luminosity relations (e.g.  Bromm \& Loeb 2007); however, it has been widely rejected for a variety of strong causes (e.g.  Cabrera et al.  2007; Amati et al. 2009; Ghirlanda et al. 2009; Krimm et al.  2009; Nava et al. 2009; Xiao \& Schaefer 2009) and most recently the authors themselves recanted their previous findings (Butler et al.  2009).  Nevertheless, we are here reconsidering the cause and effects for the basic claims of satellite-to-satellite differences.

In this paper, we first start by presenting and explaining the Nakar and Piran test, which following Band \& Preece (2005) we extend by considering bursts in a plot of their $S_{bolo}$ versus $E_{peak,obs}$.  In addition, we explain why a certain amount of violators are expected, and what the observed distributions of bursts tell us to expect.  We follow this by showing gathered data from various detectors and providing a comprehensive examination of how each detector's data performs under the Nakar and Piran test.  Following this, we provide an explanation for why the vast majority of the data sets have too many violators of the Amati limit, and therefore the Amati relation is not good as a luminosity relation.  Finally, we examine several sources of systematic offsets that are actually the cause of the Amati relation in the first place, which only further condemns the Amati relation's usefulness.

\section{The $S_{bolo} - E_{peak}$ Diagram and the Amati Relation}

\begin{figure}
\begin{center}
\includegraphics[scale=0.55]{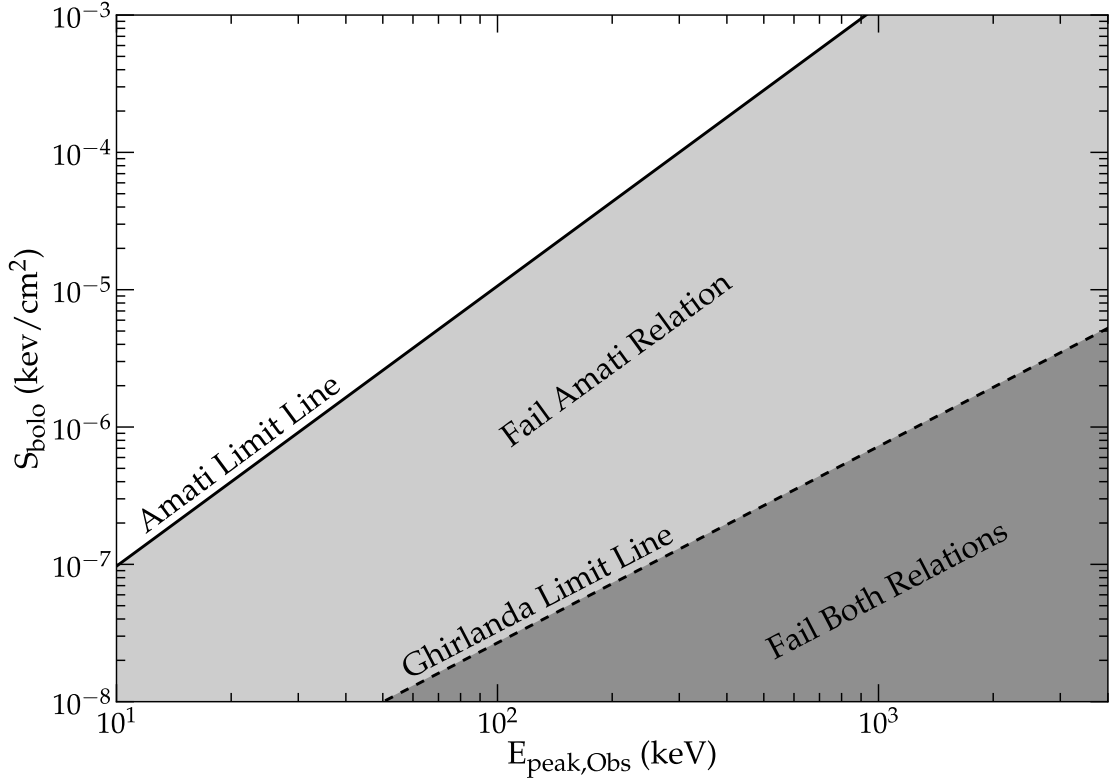}
\end{center}
\caption{The basics of the Nakar and Piran test in graphical form.  Any burst (even without a known redshift) can be plotted on this diagram.  If the Amati relation is correct, then any burst must lie above the solid line (from equation \ref{eq:AmatiLimit}), although normal scatter from measurement error will put somewhat less than half of the bursts just below the limit line.  If the Ghirlanda relation is correct, then any burst must lie above the dashed line (from equation \ref{eq:GhirLimit}), although normal scatter from measurement error can put a small fraction of the bursts just below the limit line.  If a burst lies below one of the lines, then it is called a `violator' of that relation.}
\label{fig:NaPZONES}
\end{figure}

One way to visualize bursts under the Nakar and Piran test is to plot them on a $S_{bolo} - E_{peak,obs}$ diagram.  In doing this, we can not only easily see how a certain group of bursts fares on the Nakar and Piran test, but we also determine if there is a systematic offset between different detectors.  As an example, we can determine whether different detectors are pre-disposed towards different regions on the diagram.  In addition to plotting points for individual bursts, we can also plot the Amati limit (equation \ref{eq:AmatiLimit}) and the Ghirlanda limit (equation \ref{eq:GhirLimit}) for easier visualization of where the limits lie.  Figure \ref{fig:NaPZONES} shows the basic idea behind the plots, with three zones for whether the burst violates no limit, the Amati limit only, or both limits.  

\begin{figure}
\begin{center}
\includegraphics[scale=0.55]{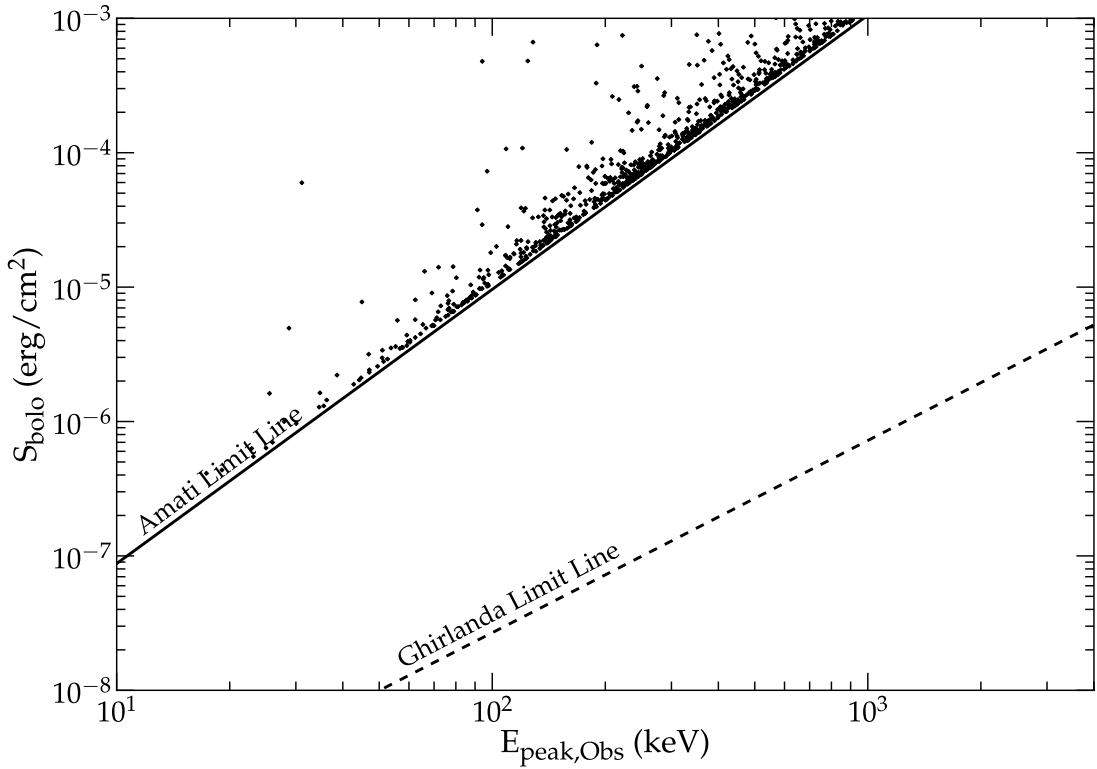}
\end{center}
\caption{1,000 simulated bursts based on the Amati relation with no measurement errors.  We assumed that the Amati relation is exact.  In our Monte Carlo simulation, each burst had a redshift chosen randomly from a reasonable cosmological model for bursts, an $E_{peak,obs}$ value chosen randomly from the log-normal distribution of Mallozzi et al. (1995), the burst energy calculated from the Amati relation, then the observed $S_{bolo}$ calculated from the burst energy and redshift.  The simulated bursts are usually close to the Amati limit line, and there are no violators.}
\label{fig:NaPAmatiMC}
\end{figure}

To illustrate the Amati limit, Figure \ref{fig:NaPAmatiMC} presents our Monte Carlo simulation of 1000 bursts where the Amati relation is adopted. There are no measurement errors, no selection effects for satellite detectors, and the burst luminosity and distance distributions are a reasonable model of the real Universe.  Our simulation for each burst starts with the random selection of $E_{peak,obs}$ as based on a log-normal distribution like in Mallozzi et al. (1995).  In addition, the redshift of the burst is randomly selected from a log normal distribution centered at \textit{z}$=$2 with a standard deviation of 1. While this is not \textit{exactly} the same as the distributions seen in cosmological models, it is similar enough, and the exact shape does not effect the conclusion. With these two values, we first find the intrinsic $E_{peak}$ by simply applying the redshift correction: $E_{peak}=E_{peak,obs}(1+z)$.  We then use the Amati relation to derive $E_{\gamma, iso}$, and use equation 2 to get the observed $S_{bolo}$.  As such, the figure shows a realistic distribution, or at least for no measurement uncertainties.  In the figure, we see that there are no violators (i.e., bursts appearing below the Amati limit), with most bursts appearing close to the limit line.  This figure is a central illustration of the Nakar \& Piran test, which we will extend in this paper.

\begin{figure}
\begin{center}
\includegraphics[scale=0.55]{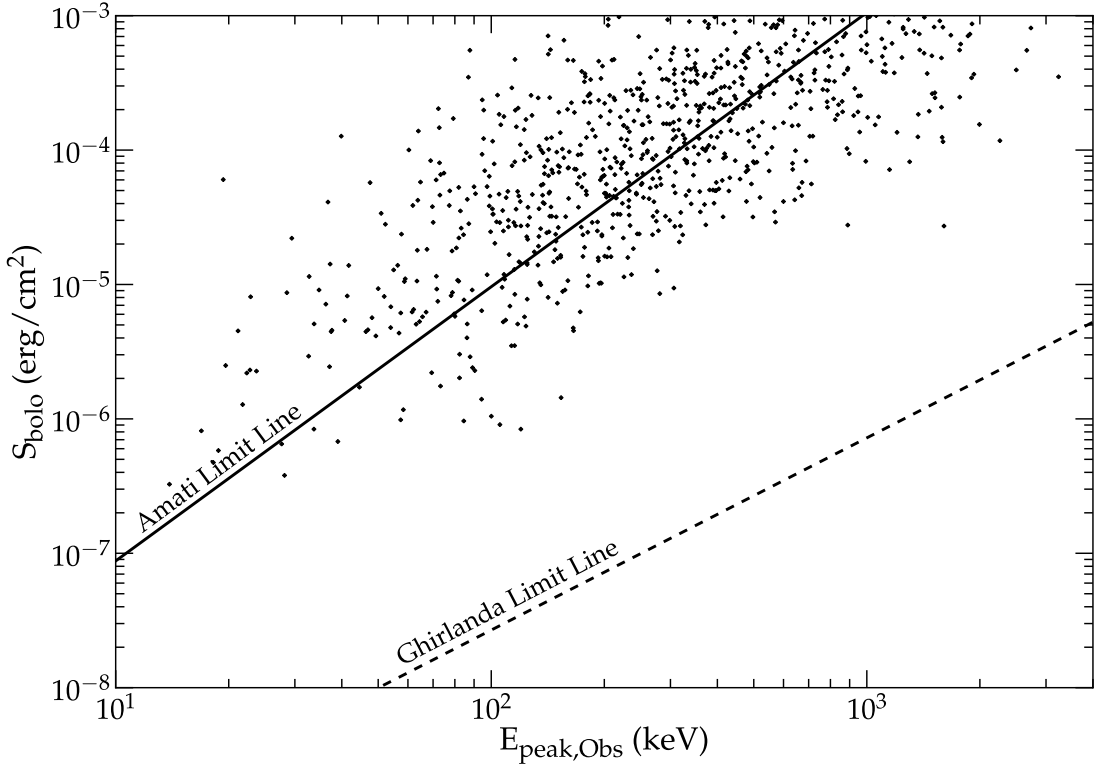}
\end{center}
\caption{1,000 simulated bursts based on the Amati relation with realistic measurement errors.  These 1000 bursts are identical with the bursts in the previous figure, except that a statistical scatter has been added to the intrinsic values.  For this Monte Carlo simulation, the measurement error is taken from a log-normal distribution where the one-sigma scatter in $\log$($E_{peak,obs}$) is 0.25 (c.f., Collazzi et al. 2011).  With this, the fraction of burst that violate the Amati limit rises from zero to $\sim$40\%.}
\label{fig:NaPAmatiMCScatter}
\end{figure}

If we allow for ordinary scatter caused by measurement errors in $E_{peak,obs}$ and $S_{bolo}$, then the tight scatter in Figure \ref{fig:NaPAmatiMC}  is lost.  This is shown in Figure \ref{fig:NaPAmatiMCScatter}, where suddenly somewhat less than half of the bursts become violators.  For this simulation, we assumed that the measurement errors have a log-normal distribution with a one-sigma width of 0.25 (Collazzi et al. 2011).  The exact fraction of violators will depend on the size of the observational scatter.  In this realistic simulation, $\sim$40\% of the bursts are below the Amati limit line.  The point of this figure is that normal and expected observational measurement errors will lead to nearly half the bursts being apparent violators.  Importantly, this scatter does not explain the high violator rates reported by Band \& Preece (2005) and Goldstein et al. (2010).  This discrepancy is the main topic of this paper.

\begin{figure}
\begin{center}
\includegraphics[scale=0.55]{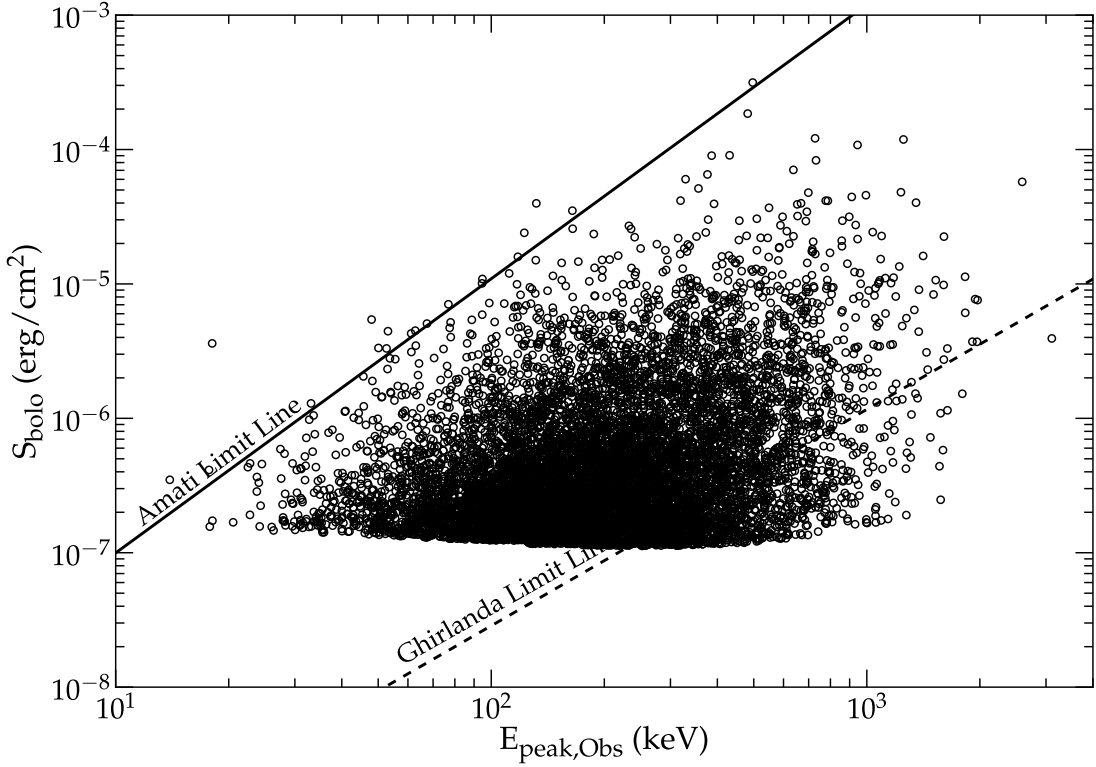}
\end{center}
\caption{10,000 simulated bursts without the Amati relation.   We start by generating a flux in the 100-500 keV range using the BATSE $\log[N(>P)] - \log[P]$ relation (Fenimore et al. 1993, Fishman and Meegan, 1995).  We then generate an $E_{peak}$ based on a log normal distribution, with a dependence on the brightness of the burst (as seen in Mallozzi et al. 1995). Finally, we use the generated $E_{peak}$ to apply an appropriate bolometric correction based on the band function with ($\alpha=-1.0$ and $\beta=-2.0$). This bolometric correction ranges from a factor of $\sim$3.5 to $\sim$7.1. The result is an \textit{illustration} of the intrinsic distribution of the population of bursts on the sky.  The point of this Figure is that the distribution covers a large area without the Amati relation.}
\label{fig:NaPDistMC}
\end{figure}

For comparison, we can also consider how the $S_{bolo} - E_{peak,obs}$ diagram would look if neither the Amati or Ghirlanda relations were valid.  For this, we have constructed another Monte Carlo simulation (see Figure  \ref{fig:NaPDistMC}).  As in Figure 2, we have assumed no measurement errors, no selection by satellite detectors, and we have adopted realistic luminosity and distance distributions, but we have made no constraints from either the Amati or Ghirlanda relations. We start by selecting burst distances and energies in the 100-500 keV such that they reproduce the observed $\log(N)-\log(P)$ curves for {\it BATSE} (Fenimore et al. 1993, Fishman \& Meegan 1995). We then generate $E_{peak}$ based on a log normal distribution with some loose connection to the brightness of the burst (as seen in Mallozzi et al. 1995). We then apply a bolometric correction with ($\alpha=-1.0$ and $\beta=-2.0$). The result is an in illustration of the intrinsic distribution of bursts on the sky.  Our simulation of 10,000 bursts has approximate edges at 20 and 3000 keV, plus lower and upper edges simply where we cut off the $\log(N)-\log(P)$ curve. The key point is that Figures 2 and 4 are greatly different, because low-fluence bursts will dominate unless some law/correlation forces these low-fluence events to have low-$E_{peak,obs}$.  So we have two extreme cases that produce greatly different distributions in the $S_{bolo} - E_{peak,obs}$ diagram.

Both Figures 2 and 4 are for the intrinsic distributions of GRBs in a realistic case with no effects of detector thresholds or measurement uncertainties.  From a comparison of Figures 2 and 3, we see that realistic measurement errors will substantially smear the underlying distribution.  Detector thresholds will also force a fuzzy cutoff roughly running along some horizontal curve.  For a detector with a high threshold, the many violators in Figure 4 will never be detected and the violator fraction might appear acceptable.  For a detector with a low threshold, we should be able to easily determine whether the Amati relation is valid.

\section{Generalizing the Test to Many Detectors}

So far, the Nakar \& Piran test has only been applied to BATSE bursts (Nakar \& Piran 2005, Band \& Preece 2005; Goldstein et al. 2010) and to a collection of bursts with redshifts as detected by a range of many satellites (Amati 2002; 2006; Schaefer \& Collazzi 2007).  But this test can be extended to many satellites, because all that is needed are values of $S_{bolo}$ and $E_{peak,obs}$, with both of these being commonly reported for many bursts.  The essence is to find the fraction of violators, $\xi$, for each sample.  We will also keep track of the quantity $\langle\log\frac{E^{2.04}_{peak}}{S_{bolo}}\rangle$ for each sample, as this can be directly compared to $1.13 \times 10^9$ (in units of $\unit{keV^{2.04} \; erg^{-1} \; cm^{2}}$) so as to test the Amati limit (cf. equation 4).  We will keep track of these statistics for both bursts with spectroscopically determined redshifts (Greiner 2010) and those with no known redshift.  The samples and their statistics are presented in Table 1, with some discussion in Sections 3.1 to 3.8.  The $S_{bolo} - E_{peak,obs}$ diagrams for each sample are presented in Figures 5-13.

\subsection{Amati et al.  (2006)}

\begin{figure}
\begin{center}
\includegraphics[scale=0.55]{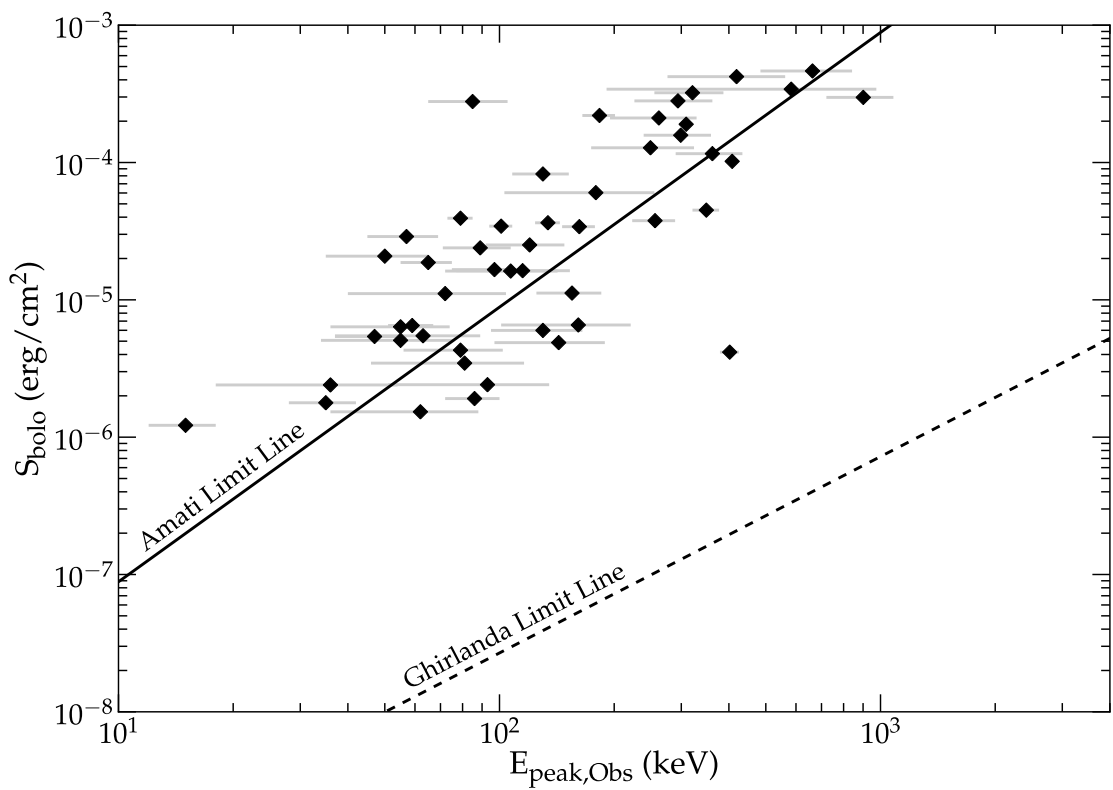}
\end{center}
\caption{The Nakar and Piran test for 50 bursts from Amati et al.  (2006).  This data came from a variety of different detectors, including \textit{Beppo}-SAX, Konus,BATSE, HETE, and \textit{Swift}.  All bursts in this sample have known associated spectroscopic redshifts.  Of the 50 bursts, 34\% fail the test.  This is within the expected failure rate of the Amati relation.}
\label{fig:NaPAmati}
\end{figure}

We start by using the compilation of data from Tables 1 and 2 of Amati et al.  (2006).  These bursts all have redshifts and are from \textit{Beppo}-SAX Konus, HETE, BATSE, and \textit{Swift}.  Using the Amati relation, we calculate the $S_{bolo}$ from the given data.  We exclude bursts 050315, 050824, 050904, 981226, 000214 and 030723 because only limits to $E_{peak,obs}$ are provided, and therefore are not useful to us.  We also had to exclude burst 980329 because a redshift range is given, and we could therefore could not get a accurate measurement of ($d_{L}$) for converting $E_{iso}$ into $S_{bolo}$. The results are in Figure \ref{fig:NaPAmati}.

This sample of GRBs was largely the same as used by Amati (2002) to discover and calibrate the Amati relation, so it is no surprise that the bursts are spread out along the Amati limit line.  The violator fraction is $\xi=34\%$, which is as expected given the usual scatter due to measurement errors.  The sample has $\langle\log\frac{E^{2.04}_{peak}}{S_{bolo}}\rangle=8.90$ which is close to the limit of $\log (1.13\times 10^9)=9.05$ (in appropriate units).  The RMS scatter of $\log\frac{E^{2.04}_{peak}}{S_{bolo}}$ is 0.56, which is a measure of how tight the Amati relation is for the sample.

\subsection{Schaefer (2007)}

\begin{figure}
\begin{center}
\includegraphics[scale=0.55]{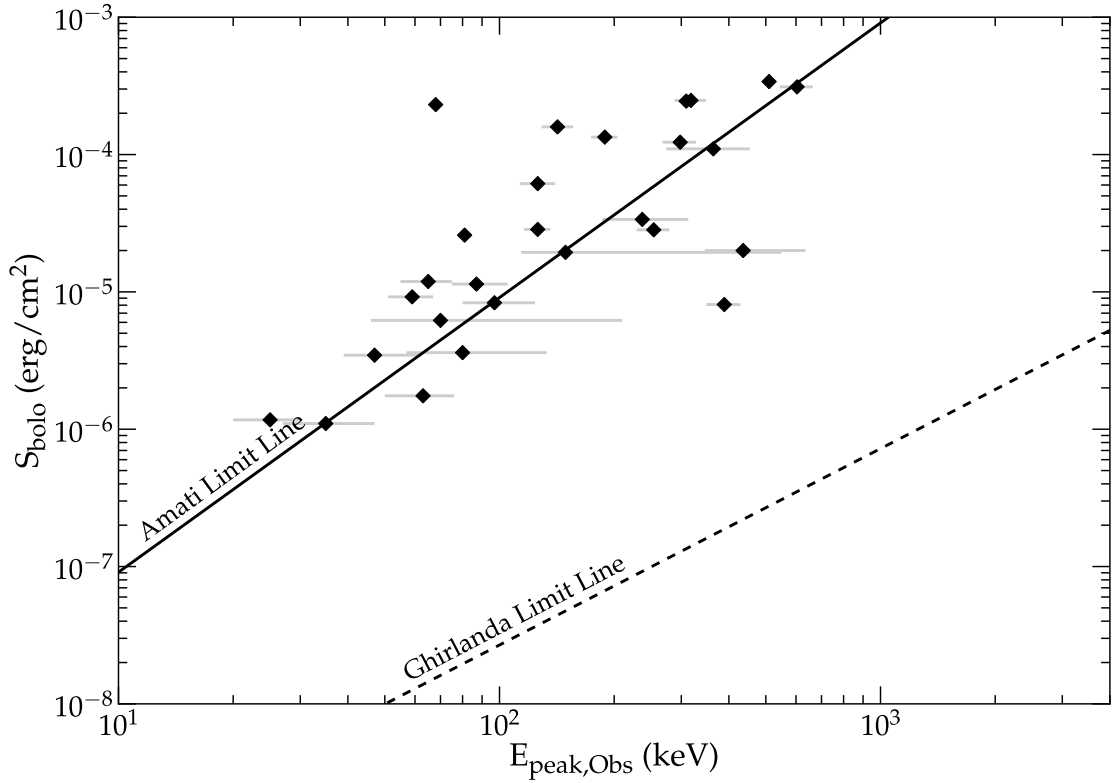}
\end{center}
\caption{The Nakar and Piran test for 27 bursts from Schaefer (2007).  This data came from a variety of different detectors, including \textit{Beppo}-SAX, Konus,BATSE, HETE, and \textit{Swift}.  All bursts in this ample have known associated spectroscopic redshifts.  Of the 27 bursts, 41\% fail the test.  This is in within the expected failure rate of the Amati relation and in agreement with previous tests on this data (see Schaefer and Collazzi 2007).}
\label{fig:NaPSchaefer}
\end{figure}

Next, we visit another compilation set, this time from Schaefer (2007).  This data set also takes its burst sample from a variety of different detectors: Konus, \textit{BATSE}, \textit{Beppo}-SAX, HETE, and \textit{Swift}  bursts were included for this sample.  While the paper studies 69 GRBs with known redshift, only 27 have the bolometric fluence reported and thus are the ones that we use (see Figure \ref{fig:NaPSchaefer}).

Of the 27 bursts, 11 fail the Nakar and Piran test ($\xi$=41\%).  This is an expected failure rate for the Amati relation and is in agreement with previous analysis on this data (see Schaefer and Collazzi 2007).  The average value for the log of $\langle\log\frac{E^{2.04}_{peak}}{S_{bolo}}\rangle$ is 8.95$\pm$0.57.  So this data is similar to the previous data set (which is not surprising as they share some of the same bursts).

The two samples which contain exclusively bursts with known redshifts both agree well with the Amati relation, and this has long been the primary justification for accepting the Amati relation as a physical relation for GRBs.  However, other samples (see below) do not agree with the Amati limit, and this suggests that bursts with redshifts might be a significantly different sample from those without redshift.  This is the reason why we distinguish bursts with and without redshifts in Table 1 and in the Figures.

\subsection{BATSE data}

\begin{figure}
\begin{center}
\includegraphics[scale=0.55]{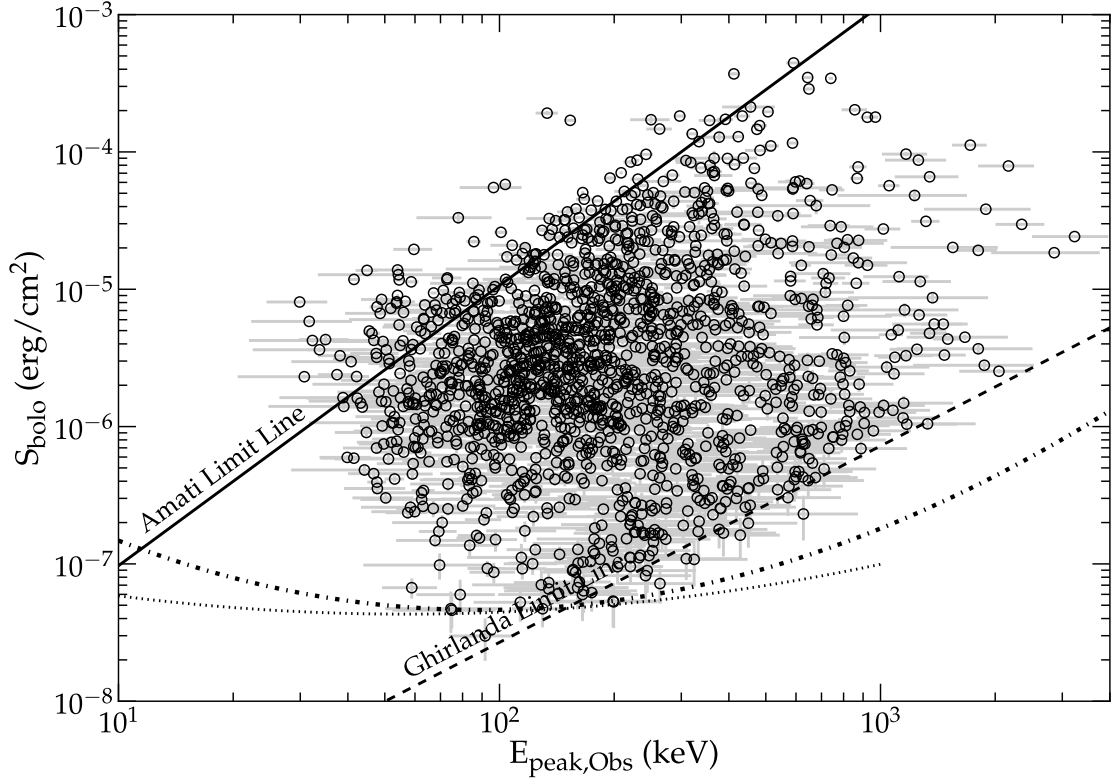}
\end{center}
\caption{1654 BATSE bursts from the future 5B BATSE catalog (Goldstein et al. 2011).  The $E_{peak}$ and $S_{bolo}$ data come from the best fit of either a CPL or Band model, whichever was significantly better.  In addition, we use only the bursts for which the relative error on the measurements are 40\% or better.  This selection fails at a very high rate, with 93\% violators.  The zone covered by the BATSE sample has a very large coverage area, but still only a very few bursts are passing.  This is particularly condemning, as it hints that any future Amati relation will also fail for the BATSE sample. The dotted line represents an \textit{illustrative} line for the trigger threshold and the dot-dashed line represents an \textit{illustrative} model of the $E_{peak}$ detection threshold (see section 4). }
\label{fig:NaPBATSE}
\end{figure}

Our data for BATSE is a part of the upcoming 5B catalog (Goldstein et al. 2011).  We present the values of $E_{peak}$ and $S_{bolo}$ for the most statistically preferred fitting model, CPL or Band.  We also use bursts for which we have 40\% relative errors or better.  After applying these selection criteria, we are left with 1654 bursts, which are plotted in Figure \ref{fig:NaPBATSE}. In the figure, we also introduce two new curved lines which represent \textit{illustrative thresholds} for the trigger (dotted) and the ability to detect $E_{peak}$ (dot-dashed). These are explained in detail in section 4. 

We find that the BATSE bursts fail at an extreme rate, with 93\% violators.  In addition, the BATSE bursts cover a large region of the disallowed zone, with very few bursts above the limit.  We find that $\langle\log\frac{E^{2.04}_{peak}}{S_{bolo}}\rangle$ to have a value of 10.18$\pm$0.88.  The failure rate is consistent with that observed in the past of BATSE bursts in previous works.  The spread of BATSE bursts is so large, it hints that most any future changes to the Amati relation (e.g. as more \textit{Swift} bursts are detected with redshifts) will result in a high failure rate.
 
\subsection{HETE data}

\begin{figure}
\begin{center}
\includegraphics[scale=0.55]{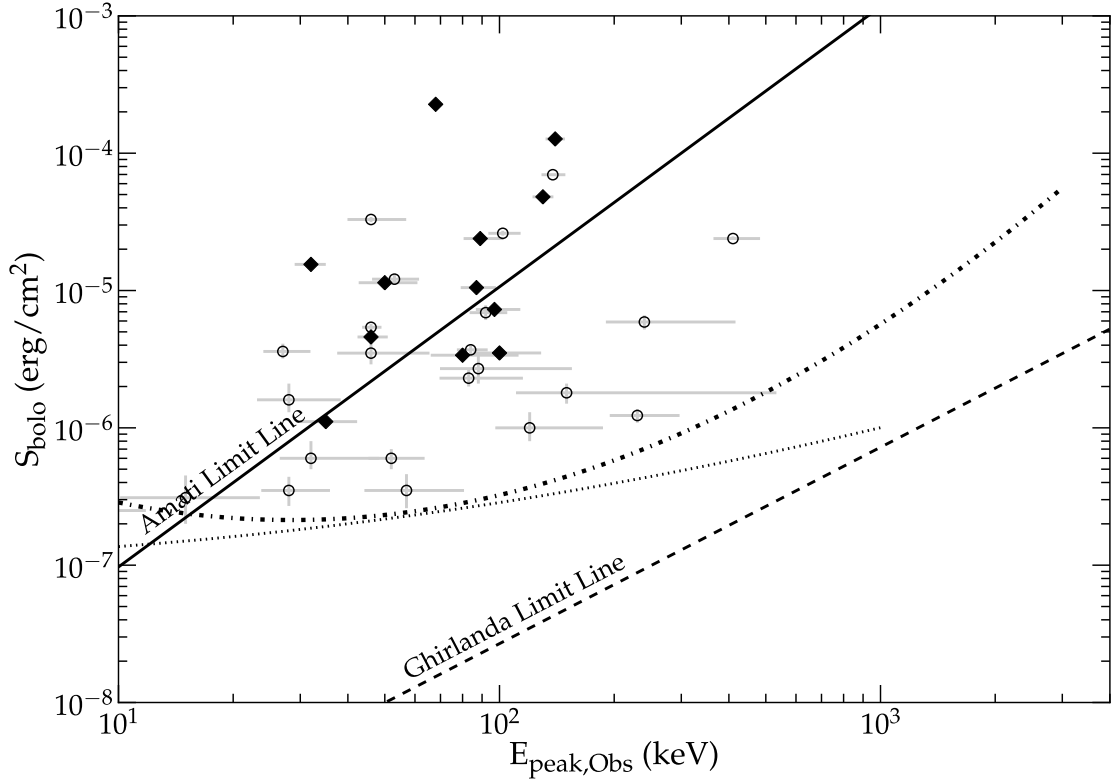}
\end{center}
\caption{HETE data from Sakamoto et al.  (2005).  In total, 44\% of the 36 bursts fail below the Amati limit line, which is within the expected failure rate.  Likewise, 33\% of the 12 bursts with associated spectroscopic redshift re violators (which again, is within the expected failure rate).  The bursts with redshift do not appear to be significantly different from those without.  The HETE data is unique in that it seems to most resemble the original data from Amati, even though the scatter around the Amati limit line is very large. The dotted line represents an \textit{illustrative} line for the trigger threshold and the dot-dashed line represents an \textit{illustrative} model of the $E_{peak}$ detection threshold (see section 4).  Filled diamonds represent bursts for which there is a measured redshift, unfilled circles are bursts for which there is no redshift.}

\label{fig:NaPHETE}
\end{figure}

Our sample of HETE bursts comes from Sakamoto et al.  (2005).  The quoted values for $S_{bolo}$ were covering the 2-400 keV range, and had to be converted into bolometric fluences.  This was done by using the given parameters for the spectral model (Band or Cut-off Power Law (CPL)) to extrapolate a bolometric correction.  This was generally a small correction, while even the large corrections are still small compared to scatter in Figure \ref{fig:NaPHETE}.  The quoted error bars for both $E_{peak}$ and $S_{bolo}$ are given for the 90\% level, so they had to be converted into standard one sigma confidence level error bars. The figure also has \textit{illustrative} lines to represent trigger thresholds (dotted line) and the $E_{peak}$ detection threshold (dot-dashed line). Again, these will be explained in detail in section 4. 

The HETE bursts have $\xi=33\%$ and $\langle\log\frac{E^{2.04}_{peak}}{S_{bolo}}\rangle$ similar to that of the original Amati sample, so it appears that these bursts are consistent with the Amati relation.  Nevertheless, the scatter apparent in Figure \ref{fig:NaPHETE} is so large that there is little utility in applying the Amati relation to these bursts.  The difference is insignificant for $\langle\log\frac{E^{2.04}_{peak}}{S_{bolo}}\rangle$ between bursts with and without redshifts.  Of all the single-satellite data sets that we consider, the HETE data is the only one that apparently obeys the Amati relation, although its large scatter limits its usefulness.

\subsection{\textit{Swift}}

\begin{figure}
\begin{center}
\includegraphics[scale=0.55]{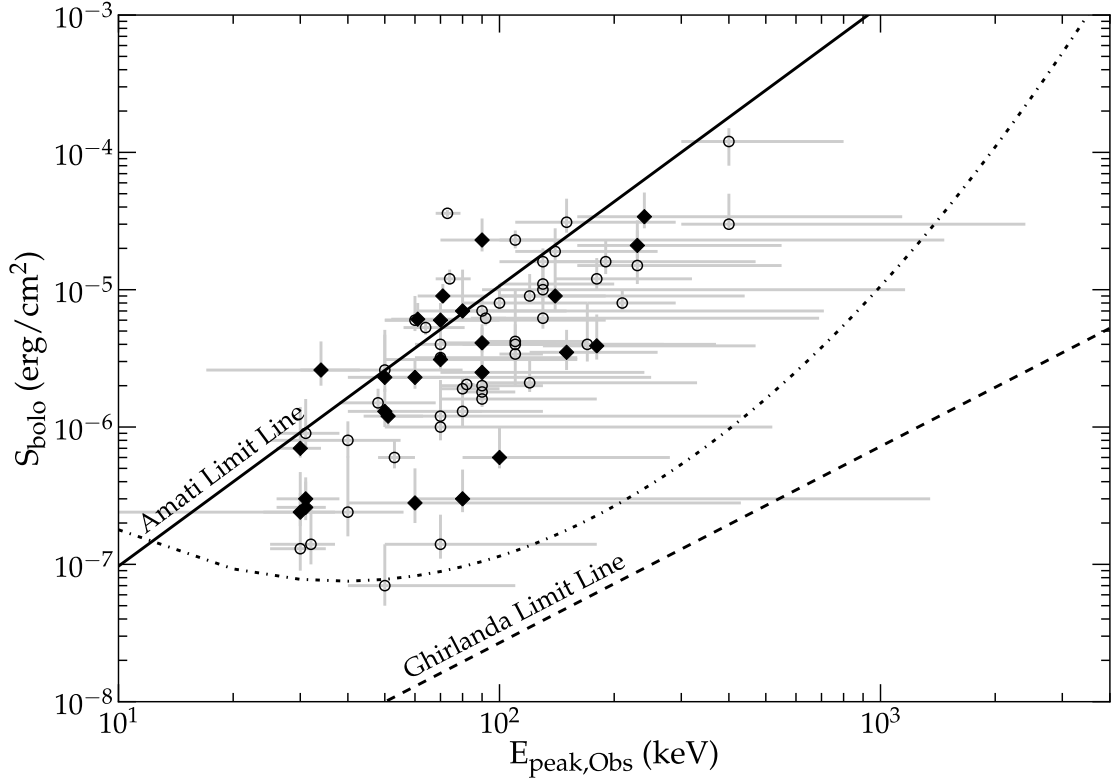}
\end{center}
\caption{\textit{Swift} data from Butler et al.  (2007).  In total, there are 71 bursts, 82\% of which violate the Amati limit.  This is far beyond the expected value, and thus the Amati relation fails for the \textit{Swift} data.  The same conclusion is reached when looking just at the bursts with known spectroscopic data, with 76\% of those 25 bursts being violators.The bursts with known redshift are not different from those without known redshift (see Table 1). The dotted line represents an \textit{illustrative} line for the trigger threshold and the dot-dashed line represents an \textit{illustrative} model of the $E_{peak}$ detection threshold (see section 4).  Filled diamonds represent bursts for which there is a measured redshift, unfilled circles are bursts for which there is no redshift.}
\label{fig:NaPSwift}
\end{figure}

For the \textit{Swift} data, we use the catalog in Butler et al.  (2007).  We use their $E_{peak,obs}$ as derived from frequentist statistics as it is the most common approach to finding $E_{peak}$.  (Their Bayesian values were made with unreasonable priors that significantly skew the results.)  The \textit{Swift} burst detector only goes up to ~150 keV, so the reported values of $E_{peak,obs}$ are almost all lower than 200 keV.  We have adopted their bolometric fluences and we have converted their non-standard 90\% error bars into standard one-sigma error bars. Figure \ref{fig:NaPSwift} plots the results; bursts without known redshifts are represented as empty circles and bursts with known redshift are represented by a filled diamond. This will be true for all future plots.  This is the last of the three plots in which we have plotted \textit{illustrative} lines representing trigger (dashed line) and $E_{peak}$ detection thresholds (dot-dashed line), which again, are detailed in section 4.

The \textit{Swift} bursts violate the Amati limit at a rate of 76\% to 82\%.  That is, the Amati relation does not work for \textit{Swift}.  This result is not the result of small number statistics, and we can see from the distribution that the disagreement is highly significant.  This is another version of the same conclusion first reported by Butler et al. (2007).

When first confronted with the discrepancy that bursts with redshifts agreed with the Amati relation (see Figures \ref{fig:NaPAmati} and \ref{fig:NaPSchaefer}) while bursts without redshifts disagreed with the Amati relation (Band \& Preece 2005), our initial thought was that the bursts with redshifts might be somehow selected from a separate population for which the Amati relation applied.  However, with this large sample of well-measured bursts from \textit{Swift}, the distributions of bursts with and without redshifts is essentially identical.  Thus, we have a proof that the success or failure of the Amati relation does not depend on some selection effect that correlates with the measuring of spectroscopic redshifts. Another thing to remember is that since \textit{Swift} bursts are the bursts that account for a majority of the bursts with known redshifts, there is a built-in selection effect that will eventually develop that will bias future iterations of the Amati limit towards the area \textit{Swift} bursts cover.

\subsection{Suzaku}

\begin{figure}
\begin{center}
\includegraphics[scale=0.55]{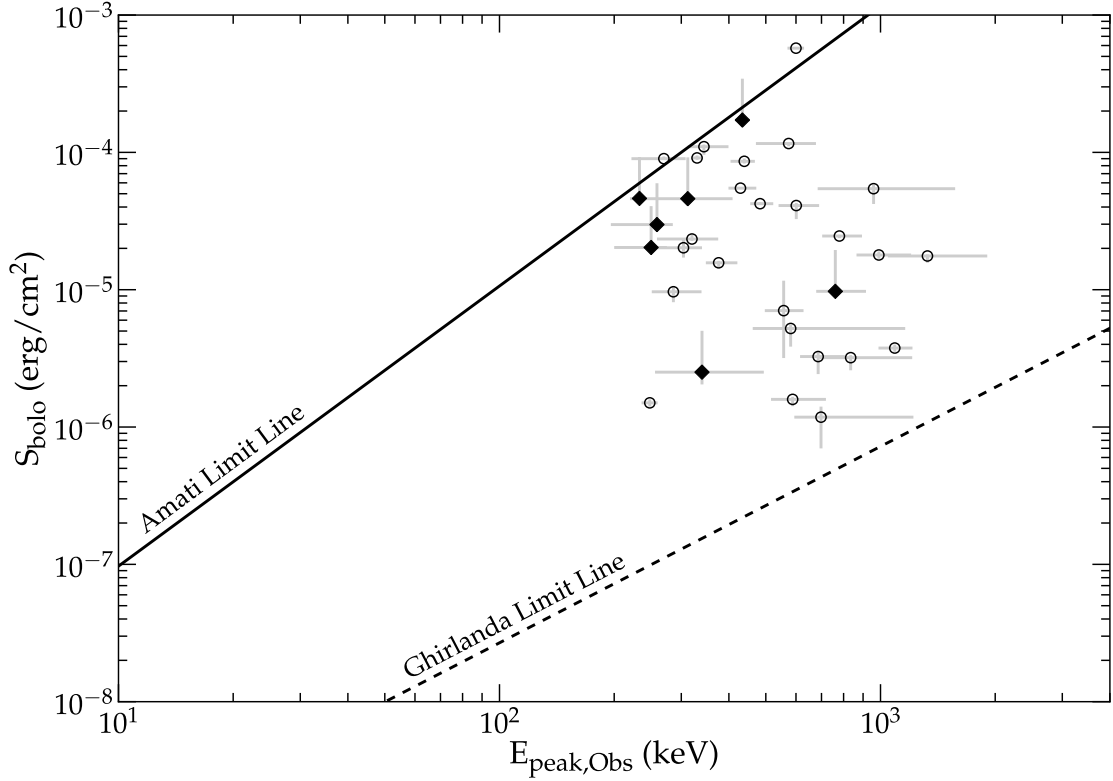}
\end{center}
\caption{Suzaku data.  We use only the bursts for which the time-integrated $E_{peak}$ are reported.  The fluences are also reported in these notices, typically over the 100 keV to 1 MeV range.  For these fluences, we had to convert them to bolometric fluences using the provided spectral parameters.  Of the 32 bursts we use, 94\% are violators of the Amati limit.  The 7 bursts that have associated spectroscopic redshift have a 100\% violator rate.  Again, the bursts with known redshift are not different from the overall sample.  Filled diamonds represent bursts for which there is a measured redshift, unfilled circles are bursts for which there is no redshift.}
\label{fig:NaPSuzaku}
\end{figure}

Suzaku data for long GRBs are available through the GCN circulars (http://gcn.gsfc.nasa.gov).  Typically, the reported fluence covers the 100 keV to the 1 MeV range, so we apply a bolometric correction based on the reported spectral fit.  A typical bolometric correction value is a factor of $\sim$1.7.  The $E_{peak,obs}$ and fluence values reported in the circulars are preliminary and made soon after the burst, yet any likely changes to get to the final best fits are greatly smaller than the scatter shown in Figure  \ref{fig:NaPSuzaku} and are thus not important.  The Suzaku bursts all have $E_{peak,obs}>200$ keV, a result of the relatively high energy range of sensitivity of the detector.

Most of the Suzaku bursts violate the Amati limit, usually by a large factor ($\xi \sim, 94\%$) and the small fraction that are not violators are very close to the limit (Figure  \ref{fig:NaPSuzaku}).  Therefore, the Amati relation does not work for Suzaku bursts.  This is true for both bursts with redshifts and without.

\subsection{\textit{Swift}-Suzaku}

Krimm et al. (2009)  presented a catalog of bursts for which there was both \textit{Swift} and Suzaku data.  The expanded energy range gave better fits to the spectra, with \textit{Swift} covering the lower energies and Suzaku covering the higher energies.  With the joint spectral fits over a very wide range of photon energies, the  sample has a wide range of $E_{peak,obs}$ values from 30 keV to 2000 keV.  The catalog lists the best fit for the three major spectral models (power law, power law with an exponential cutoff, and Band model) for the majority of the listed bursts, while we use the $E_{peak,obs}$ values from just the Band function.

\begin{figure}
\begin{center}
\includegraphics[scale=0.55]{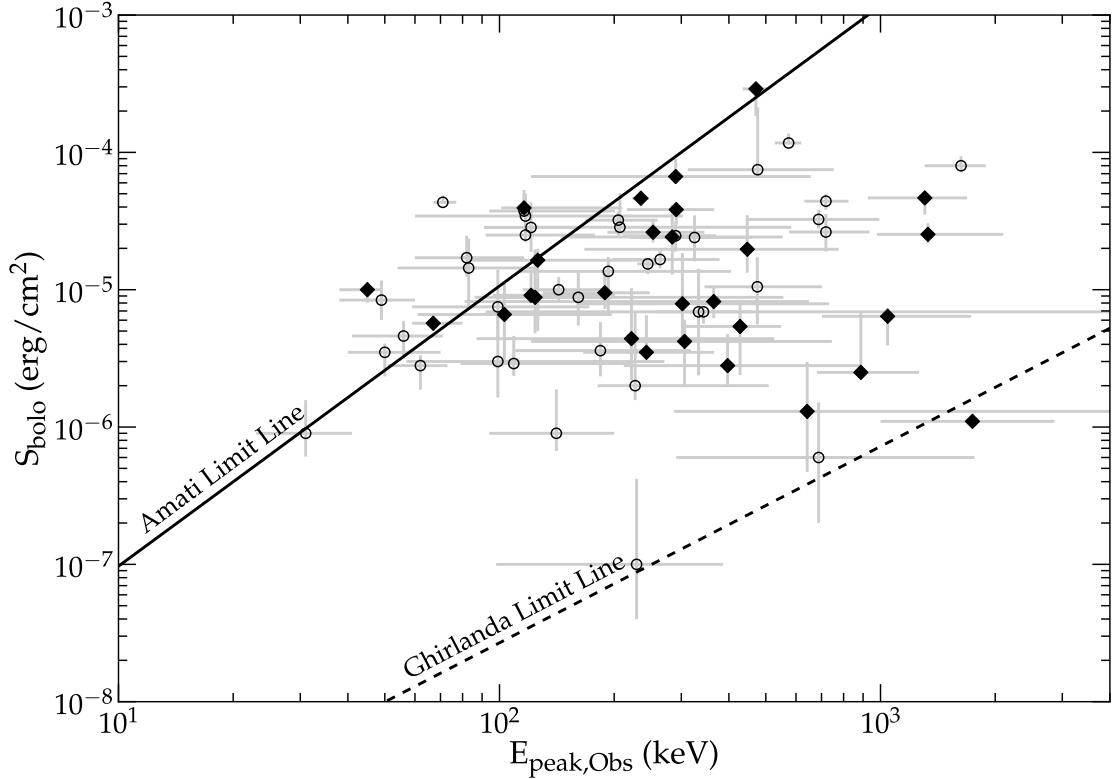}
\end{center}
\caption{Combined data from \textit{Swift} and Suzaku.  Krimm et al.  (2009) took the raw data from both detectors and fit the combined spectra to get a better measurement of $E_{peak}$ for a large sample of bursts.  We use their $E_{peak}$s as found from the Band function.  38 of the 66 usable bursts do not have spectroscopic redshifts, of which 86\% are violators.  These bursts have an average log of the energy ratio of 9.63$\pm$0.85, whereas the Amati relation requires that this average must be less than 9.04.  28 of the bursts have spectroscopic redshifts, 86\% of those busts are violators, with an average log of the energy ratio of 10.01$\pm$1.01.  Even with the broad spectral range provided by combining the \textit{Swift} and Suzaku data, the Amati relation fails.  Filled diamonds represent bursts for which there is a measured redshift, unfilled circles are bursts for which there is no redshift.}
\label{fig:NaPSwiftSuz}
\end{figure}

The joint sample largely stretches between the Amati limit line and the Ghirlanda limit line (Figure \ref{fig:NaPSwiftSuz}).  The fraction of violators is 86\% for the 28 bursts with spectroscopic redshifts and 74\% for the 38 bursts without redshifts.  Again, the Amati relation fails, and there is no significant difference related to whether the burst has a spectroscopic redshift or not.  No burst significantly violates the Ghirlanda limit.

\subsection{Konus}

\begin{figure}
\begin{center}
\includegraphics[scale=0.55]{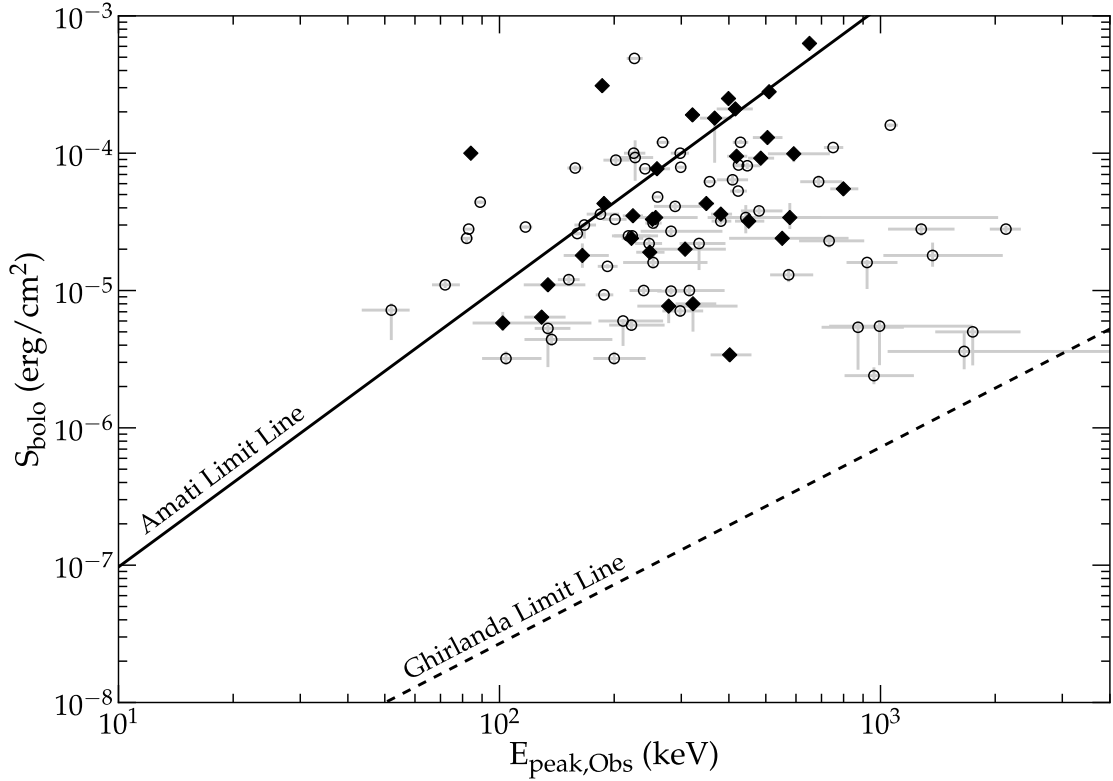}
\end{center}
\caption{Konus bursts.  This distribution of bursts has a fairly flat bottom corresponding to the trigger threshold for the detector.  The very broad energy range of Konus allows for $E_{peak,obs}$ values to be measures from 30 keV to 2000 keV.  The fraction of bursts violating the Amati limit is 73\% for the 33 bursts with spectroscopic redshifts and 78\% for the the 64 bursts without redshifts.  With bursts extending down to near the Ghirlanda limit line, the observed distribution is clearly not that of the Amati relation plus some ordinary measurement errors.  In other words, the Amati relation fails for this sample. Filled diamonds represent bursts for which there is a measured redshift, unfilled circles are bursts for which there is no redshift.}
\label{fig:NaPKonus}
\end{figure}

The GRB detectors on the Wind satellite (Aptekar et al., 1995) are long-running instruments with a stable background that has measured many bursts, with the fluence and $E_{peak,obs}$ values promptly reported in the GCN circulars.  These reported value are preliminary, with no final analysis having been published, but any plausible errors due to the preliminary nature of the report are greatly smaller than the observed scatter in the $S_{bolo} - E_{peak,obs}$ diagram (see Figure \ref{fig:NaPKonus}).  We find a total of 97 bursts, 33 of which have associated spectroscopic redshifts, with reported fluences and $E_{peak,obs}$ for the entire burst interval.  We applied a bolometric correction factor, based on the given spectral fits.  This factor was typically very small, due to the large range the reported Konus fluences usually cover.

The distribution of the Konus bursts in the $S_{bolo} - E_{peak,obs}$ diagram has a flat lower cutoff, likely due the trigger threshold (although this cutoff is higher the reported trigger threshold in Aptekar et al. (1995) of $\sim 5\times10^{-7} \unit{erg} \unit{cm}^{-2}$). The distribution also shows a fairly high upper limit on $E_{peak,obs}$ due to the sensitivity of the detectors to high photon energies.  From 22\% to 27\% of the bursts are above the Amati limit line, while all the bursts are above the Ghirlanda limit line.  The Amati relation fails for the Konus bursts.  Again, the bursts with redshifts are distributed identically to those without, so there is no apparent selection effect based on spectroscopic redshifts.

\subsection{\textit{Beppo}-SAX}

Guidorzi et al. (2010) provides a large catalog of both $E_{peak}$ and $S$ from the \textit{Beppo}-SAX GRBM.  In the catalog, data for the brightest 185 bursts are given; for which we are able to use 129.  Of the useable 129 bursts, we have only 10 bursts with spectroscopic redshifts.  The provided $S_{bolo}$ were over the 40-700 keV range.  We apply the same type of bolometric correction as before, using the provided spectral indices for the CPL used.  This correction is typically small, with a typical correction value of 1.5.

\begin{figure}
\begin{center}
\includegraphics[scale=0.55]{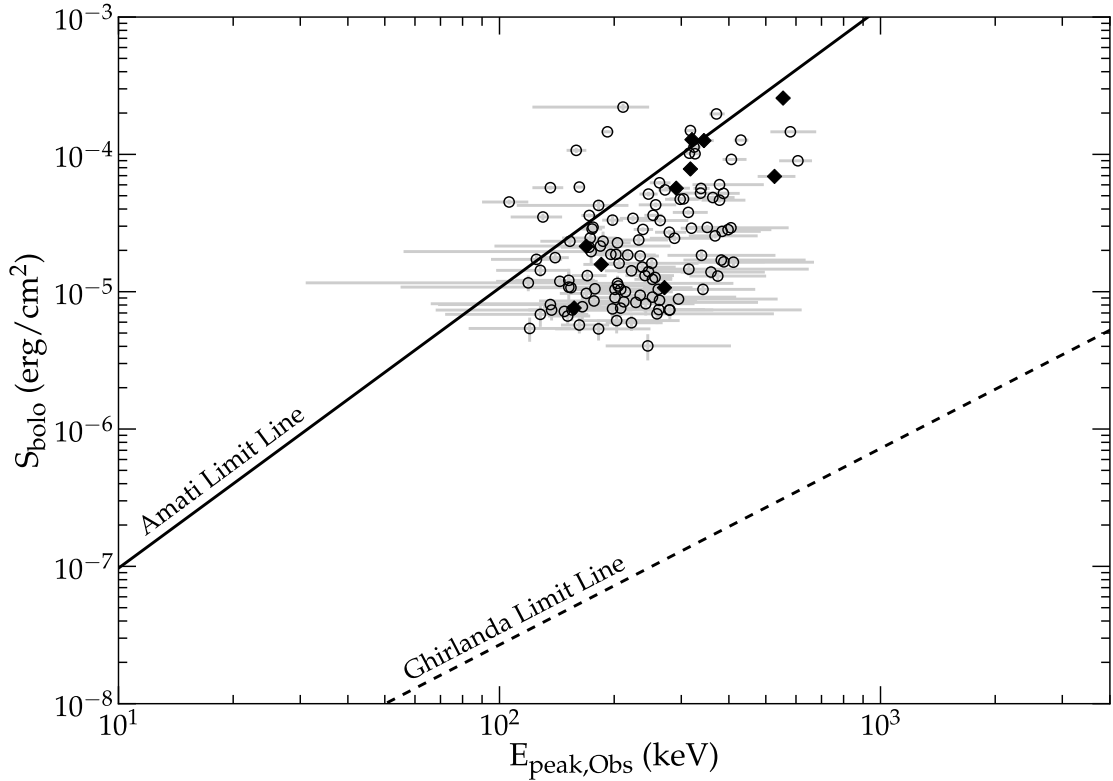}
\end{center}
\caption{\textit{Beppo}-SAX bursts.  These bursts are taken from the Guidorzi (2010) bright \textit{Beppo}-SAX burst catalog, of which we could use 119 bursts.  We find that the fraction of bursts violating the Amati relation is 85\% for bursts without spectroscopic redshifts, and 90\% for bursts with redshifts.  Because these are only the bright bursts, we cannot make any commentary as to the distribution of bursts like we do with other detectors.  We note that despite these being the brightest bursts, this sample has a high violator rate.  Since the brightest bursts are the most likely to pass the Nakar and Piran test, we can say with some confidence that the Amati relation fails for \textit{Beppo}-SAX bursts. Filled diamonds represent bursts for which there is a measured redshift, unfilled circles are bursts for which there is no redshift.}
\label{fig:NaPSAX}
\end{figure}

While it is impossible to make a strong statement about the shape of the distribution of \textit{Beppo}-SAX bursts with only the bright bursts, there is still an important result from the data.  The data is plotted in Figure \ref{fig:NaPSAX}.  We find that even among the brightest bursts, 90\% of bursts with redshifts and 90\% of bursts without redshifts are violators.  What is particularly provocative about this result is that these are the \textit{brightest} bursts, and thus the \textit{most} likely to not be violators.  It is unlikely that there are a significant number of `missing' bursts that would be non-violators.  Any such burst would have to be both bright and have a low $E_{peak}$ while still being sim enough to be missed in the bright burst catalog.  Finally, we provide information as to the average energy ratio of these bursts, but we once again stress that these are only the brightest bursts, and therefore these values should be taken with caution.  We therefore feel confident in saying that the Amati relation fails for \textit{Beppo}-SAX bursts, although this statement is not as strong as it is for other detectors because of the sample used.

\subsection{Overview of Results}

Previously, Butler et al. (2007) had pointed out that the normalization constant for the Amati relation was slightly different depending on whether \textit{Swift} or pre-\textit{Swift} bursts were used, and this is like noting that $\langle\log\frac{E^{2.04}_{peak}}{S_{bolo}}\rangle$ has changed.  Previously, Band \& Preece (2005) and Goldstein et al. (2010) pointed out that $>80$\% of BATSE bursts violate the Amati limit.  In this section, we have generalized these analyses, both to looking at many GRB detector instruments and to looking at the two dimensional distribution in the $S_{bolo} - E_{peak,obs}$ diagram.

All of these data sets give consistent conclusions:  (1) The distribution of bursts in the $S_{bolo} - E_{peak,obs}$ diagram varies significantly and greatly from satellite-to-satellite.  (2) The only data sets to pass the generalized Nakar \& Piran test for the Amati relation are the early heterogeneous sample of bursts with measured spectroscopic redshifts.  (3) The bursts detected by BATSE, \textit{Swift}, Suzaku, and Konus all have a high fraction ($\xi>70\%$) of bursts which violate the Amati limit, with the violations being highly significant and by large factors.  That is, the Amati relation fails for bursts from these four satellites.  (4) The Amati limit is satisfied for the HETE bursts, to the extent that the violator fraction is consistent the Amati relation plus normal observational scatter, however, the scatter in the $S_{bolo} - E_{peak,obs}$ diagram is so large that we conclude that the Amati relation does not satisfactorily apply to the HETE data.  (5) We find that no bursts, from any satellite, significantly violate the Ghirlanda limit.  (6) These conclusions are true whether we examine only bursts with spectroscopic redshifts or without redshifts.

The normalization factor for the Amati relation will scale closely with $\langle\log\frac{E^{2.04}_{peak}}{S_{bolo}}\rangle$.  These values are listed in Table 1, and can be taken as an intercept with some dispersion.  That is, these values can possibly be seen as a means of describing a sort of Amati relation to each detector.  However, this is not a useful interpretation as the dispersion is often as big as the population itself, rending such an interpretation nearly meaningless.  If anything, the results from Table 1 demonstrate that the Amati relation must vary from detector-to-detector by over an order of magnitude.  With the bursts seen in the sky not depending on the satellite, the large variations in the Amati relation from detector-to-detector imply that there must be some selection effect which biases the visible bursts, with these biases being instrument-specific.

Every burst detector has a substantially different distribution of bursts in Figures \ref{fig:NaPBATSE} - \ref{fig:NaPKonus}.  Since the population of bursts that appear in the skies above the Earth does not change with the satellite, so the large changes from detector-to-detector can only be due to some selection effect where bursts in various regions of the $S_{bolo} - E_{peak}$ diagram are not selected.  The next section will investigate and identify these selection effects that create the Amati relation.

\section{The Amati Relation Comes from a Combination of Selection Effects}

The distributions of bursts in the $S_{bolo} - E_{peak,obs}$ diagram are caused by a variety of effects.  Some of these effects are caused by detector limitations that prevent a burst from appearing in some parts of the $S_{bolo} - E_{peak,obs}$ diagram (Sections 4.1 and 4.2), while other effects make for rare bursts in other regions of the $S_{bolo} - E_{peak,obs}$ diagram (Sections 4.3 and 4.4).  The combination of these effects will produce the observed distributions (Section 4.5).  For some detectors, the selection effects will force the observed bursts to follow a roughly diagonal region (with wide scatter) that will appear as the Amati relation (Section 4.6).

\subsection{Trigger Thresholds}

The best known selection effect is the detector trigger threshold.  For example, a burst would trigger BATSE only if it was produced a peak flux (in a 0.064, 0.256, or 1.024 second time bin) brighter than 5.5-$\sigma$ above background in at least two detectors over the 50-300 keV energy range.  Other satellites have more complex trigger algorithms (for example, GBM has overlapping triggers), but they all come down to the same essentials.  The trigger threshold depends on the $E_{peak,obs}$, the spectral energy range of the trigger, the background flux, and the effective area of the detector.  The triggers operate off the peak flux ($P_{max}$), so the limiting fluence will depend on the effective duration ($S_{bolo}/P_{max}$), which can vary widely from burst to burst.  Thus, the limit due to trigger thresholds will be `fuzzy', with no sharp edge but rather a gradient as $S_{bolo}$ is reduced.  Approximately, the trigger threshold will produce a horizontal cutoff at the bottom of the $S_{bolo} - E_{peak,obs}$ diagram.

In principle, the exact trigger thresholds can be calculated for every detector and burst.  In practice, the conditions ($E_{peak,obs}$, background flux, incidence angle, burst light curve) vary greatly from burst to burst, creating substantial scatter in the thresholds.  For this paper, we do not need an accurate distribution for the $S_{bolo}$ threshold, so instead we calculate the typical $S_{bolo}$ threshold as a function of $E_{peak,obs}$ for average conditions.  In particular, we adopt an average spectral shape as the Band function (Band et al. 1993) with a low-energy power law index of -1.0 and a high-energy power law index of -2.0.  We also take the effective duration of the peak ($S_{bolo}/P_{max}$) such that it fits the observed distribution of the detectors.  For each detector, we take its trigger energy range, face-on effective area, and the average background flux.  The formalism and many of the input parameters were taken from Band (2003).  The result is a lower limit in the $S_{bolo} - E_{peak,obs}$ diagram, as displayed in Figure \ref{fig:NaPBATSE} for BATSE, Figure \ref{fig:NaPHETE} for HETE, and Figure \ref{fig:NaPSwift} for \textit{Swift}.  We do not have enough information to calculate trigger thresholds for some satellites, but the threshold is usually fairly obvious (e.g., Figure \ref{fig:NaPKonus} has a nearly flat and moderately sharp lower limit to $S_{bolo}$).  These thresholds are not sharp, so bursts can easily appear somewhat below the threshold.  Indeed, by varying the input conditions somewhat, the trigger threshold lines can be translated up and down substantially.

\subsection{Threshold For Measuring $E_{peak,obs}$}

A second detector selection effect is that the burst must have enough photons recorded for the analyst to be able to determine the $E_{peak,obs}$ value.  This will depend on both $S_{bolo}$ and $E_{peaks,obs}$ as well as the detector properties.  For example, a burst just above the trigger threshold will have just enough photons to be detected but not enough photons to allow any constraints on the $E_{peak,obs}$ value, so this burst will not be included in a sample for plotting on the $S_{bolo} - E_{peak,obs}$ diagram.  For another example, consider a burst with $E_{peak,obs}$ at the upper edge of the measured spectral range for a detector, such that a very bright burst will have a well-measured turnover that accurately defines the fitted $E_{peak,obs}$ value, whereas a fainter burst will have poor photon statistics near the turnover in the spectrum and the $E_{peak,obs}$ value will remain unmeasured and the burst will not be included in any of our samples.

In general, for a given $E_{peak,obs}$, there will be some lower limit on $S_{bolo}$, below which there will be too few photons to measure $E_{peak,obs}$.  As $E_{peak,obs}$ moves to higher energies, the limit on $S_{bolo}$ will sharply increase.  The result will roughly be a diagonal line across the $S_{bolo} - E_{peak,obs}$ diagram, from lower left to upper right, with any burst below that line not having a measured $E_{peak,obs}$ and not appearing in any sample of bursts in Section 3.

We have made calculations of this threshold curve for BATSE, HETE, and \textit{Swift}.  To do this, in a Monte Carlo sense, we constructed many simulated bursts over each detector's spectral range for many values of $E_{peak,obs}$ where the normalization and error bars of the spectra were determined by the burst fluence.  These spectra were also for each spectrum, we then fitted both a power law times an exponential model (with the calculated $E_{peak,obs}$ value) and a simple power law model.  If the chi-square values for the two fits differed by more than 15.0 (so that the model with the peak was a sufficiently good improvement on power law model given the extra degree of freedom), then we took the $S_{bolo}$ for the burst to be above the threshold.  By varying the $E_{peak,obs}$, we were able to determine the threshold for measurement as a curve in the $S_{bolo} - E_{peak,obs}$ diagram. As these lines are merely for illustration, we do \textit{not} use the DRM for these simulations. For BATSE, HETE, and \textit{Swift}, our calculated thresholds are presented as curves in Figures \ref{fig:NaPBATSE}, \ref{fig:NaPHETE}, and \ref{fig:NaPSwift}.  

\subsection{The $E_{peak,obs}$ Distribution}

Amongst bursts appearing in the skies, the $E_{peak,obs}$ distribution is not flat, but rather bursts appear with a roughly log-normal distribution of $E_{peak,obs}$.  For bright bursts, the mean value is 335 keV, with the FWHM stretching from roughly 150-700 keV (Mallozzi et al. 1995).  This mean value shifts significantly as the bursts get dimmer, being 175 keV just above the BATSE trigger threshold (Mallozzi et al. 1995).  The so-called `X-ray Flashes' are simply bursts in the low-energy tail of the distribution (Kippen et al. 2004; Sakamoto et al. 2005; P\'{e}langeon 2008). The existence of this single peak in the $E_{peak,obs}$ histogram is highly significant and not from any instrumental or selection effect (Brainerd et al. 1999).  In all, most bursts are between 100-700 keV, and bursts $<$30 keV or $>$1000 keV are rare.  This will directly translate to unpopulated regions of the $S_{bolo} - E_{peak,obs}$ diagram.  A direct simulation of this distribution is given in Figure \ref{fig:NaPDistMC}.

The $E_{peak,obs}$ distribution will cause definite but gradiated cutoffs in the $S_{bolo} - E_{peak,obs}$ diagram.  These cutoffs will be nearly vertical.  The drop in the average $E_{peak,obs}$ will make the cutoff on the right have a slope down to the lower left.

\subsection{The $S_{bolo}$ Distribution}

Unsurprisingly, bright bursts are rare, while faint bursts are more frequent.  The distribution of burst fluences is traditionally represented by the $\log N(>P) - \log P$ curve, for which the best observations come from the BATSE catalog (Fishman et al. 1994; Paciesas et al. 1999).  For bright bursts, the slop of the curve is nearly the ideal -$\frac{3}{2}$.  The slope flattens out for faint bursts, approaching -0.7.  In the $S_{bolo} - E_{peak,obs}$ diagram, the density of bursts falls off drastically from bottom to top (see Figure \ref{fig:NaPDistMC}).

\subsection{The Effects in Combination}

\begin{figure}
\begin{center}
\includegraphics[scale=0.55]{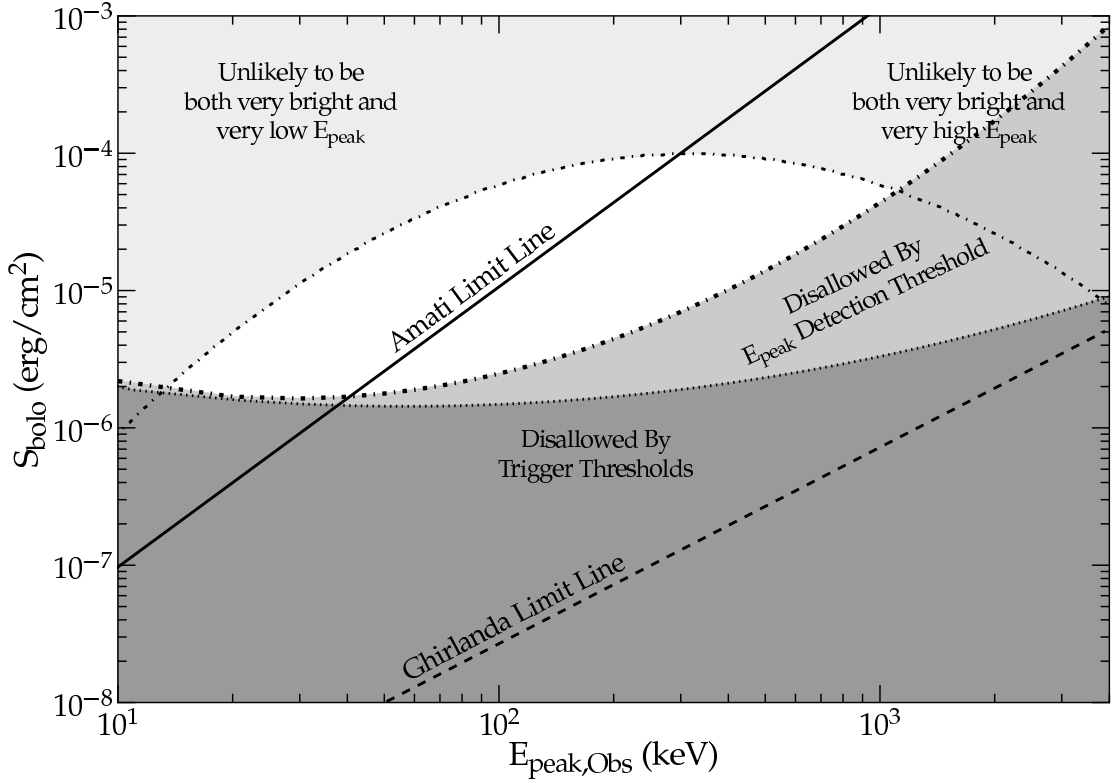}
\end{center}
\caption{The selection effects.  The two selection effects based on the intrinsic distribution of the burst population are combined and displayed as contours of burst density.  These appear as two roughly concave-down parabolas, with each representing a different density level.  The outside region is shaded darkly so as to indicate that bursts in those regions are rare, while the middle region is shaded a light gray to indicate that bursts in those areas of the diagram are less common than those in the central area.  Of the two detector selection effects, the more restrictive is the requirement that the burst be bright enough to measure $E_{peak,obs}$.  We show versions of these detector effects to illustrate how the Amati relation bursts could be seen.  The lower line illustrates a poor detector threshold (with shading below to indicate that no bursts in that area can be measured and placed onto the plot).  The other line illustrates the result of a detector with both a poor detector threshold and low energy range (with shading below it).  For a poor detector, the bursts that can be published and placed on this diagram are all in the unshaded and unhatched regions.  The point of this diagram is that the selection effects will force the plotted bursts to roughly lie along the Amati limit line, and these bursts will then appear to obey the Amati relation.  Thus, simple selection effects create the Amati relation, at least for some samples of bursts.}
\label{fig:NaPCuttoff}
\end{figure}

The intrinsic distribution of bursts in the $S_{bolo} - E_{peak,obs}$ diagram is determined by the $E_{peak,obs}$ log-normal distribution that changes with $S_{bolo}$ (Mallozzi et al. 1995) and by the $\log N(>P) - \log P$ distribution (Fishman et al. 1994).  With these two effects, the burst density across the diagram is displayed in Figure \ref{fig:NaPDistMC}.  Together, the two effects produce contour lines of burst density in the diagram, and two such curves are displayed in Figure \ref{fig:NaPCuttoff}.  The combined effects make for the upper-left corner of the $S_{bolo} - E_{peak,obs}$ diagram, simply because bursts up there are doubly rare (both low in $E_{peak,obs}$ and very bright).  This means that there are few bursts significantly above (and to the left) of the Amati limit line.  That is, there is a natural cutoff on the top side of the Amati relation.

The detector selection effects then operate on the natural distribution.  The well known trigger threshold is actually below the threshold for measuring an $E_{peak,obs}$ value, so only the later selection effect is really operating.  This selection effect cuts on a sort of diagonal from lower-left to upper-right, and its position depends greatly on the detector sensitivity and energy band for the trigger.  For a relatively poor sensitivity and a trigger energy band that effectively does not get much above a few hundred keV, the threshold will be quite high.  Indeed, for many detectors, the threshold will be just below the Amati limit line (Figure \ref{fig:NaPCuttoff}), so there will be few bursts significantly below the Amati limit line.  That is, there is a selection effect from the intrinsic distribution of bursts such that there is a natural cutoff below the Amati limit.

For some detectors, we can see that the Amati relation is a natural and expected consequence of the intrinsic burst distribution combined with normal detector selection effects.  This is illustrated in Figure \ref{fig:NaPCuttoff}, where the allowed region is confined to an area along the Amati limit line.  From Figure \ref{fig:NaPAmatiMCScatter}, we know that bursts along the Amati limit in the $S_{bolo} - E_{peak,obs}$ diagram will then imply a relation close to the Amati relation.  Thus, the natural distribution of bursts makes for bursts above the Amati limit (i.e., the very-bright low-$E_{peak,obs}$ bursts) to be rare, while the detector selection effects makes for bursts below the Amati limit (i.e., the faint high-$E_{peak,obs}$ to be too faint to have a measured $E_{peak,obs}$.  With the only bursts remaining being close to the Amati limit line, a relation like the Amati relation would be apparent.  Thus, we conclude that the Amati relation is simply a result of selection effects and there is no physical basis.

For the original bursts used to define the Amati relation (Amati 2003), an additional selection effect is operating, in that the burst must also have a measured spectroscopic redshift for inclusion in the sample used for calibration.  The selection effects for measuring a redshift are complex.  There is certainly a selection based on redshift, with the cause being that more distant bursts are fainter, hence less likely to have a visible optical transient or host galaxy bright enough to get lines in a spectrum.  An additional effect on redshift relates to the availability of spectral lines in the optical band.  The efficiency of measuring redshift as a function of redshift has been quantified in Xiao \& Schaefer (2011), with this effect being roughly an order of magnitude between $z=5$ and nearby bursts.  The efficiency of measuring redshifts also presumably depends on $S_{bolo}$ which will roughly scale as the burst brightness in the optical band.  The redshift measurement efficiency also is time dependent, as optical follow-up strategies and capabilities change within our community.  Thus, \textit{Swift} bursts started out with an average redshift of 2.8 in the first year after its launch (Jakobsson et al. 2006), while the average redshift has steadily declined to 2.1 over the last year (Jakobsson et al. 2008).  The reason for this shift is unknown, but it must come from overall follow-up practices in our community.  The bursts with redshift used in the \textit{original} calibration of the Amati relation have an average redshift of 1.5, indicating that the effective threshold for this sample is quite high.

The distribution of bursts in the $S_{bolo} - E_{peak,obs}$ diagram will depend greatly on the detector.   The threshold for measuring $E_{peak,obs}$ varies substantially detector to detector. For example BATSE has a low threshold while Konus has a high threshold.  The shape of the threshold (as a function of $E_{peak,obs}$) also varies, from a flat bottom for Konus due to its sensitivity to high energy, to the up-sloping threshold for \textit{Swift} due to its lack of high energy sensitivity and even more exaggerated in HETE with it's small area.  The ability to measure $E_{peak,obs}$ depends critically on the energy range of the spectra.  The Konus detectors have a very wide range of spectral energy resulting in a wide range of measured $E_{peak,obs}$ values, the \textit{Swift} detectors cutoff around a few hundred keV, while the Suzaku detectors can only record $E_{peak,obs} \gtrsim 200$ keV.  The combination of these selection effects makes the distribution of bursts different for each detector, and accounts for the wide range of distributions seen in Figures \ref{fig:NaPAmati} - \ref{fig:NaPKonus}.

Still, the issue has been raised in recent tests (e.g. Ghirlanda 2011) that what we are seeing in these failures is just the scatter about a relation which is ever changing with new bursts every day.  The primary argument is that the Nakar and Piran test limit should be formulated from the 3-sigma line about the model, instead of the model line itself.  As a result, the Amati limit would be considerably higher.  There are a variety of problems with this argument.  The first of which being that there is already an allowance made for the Amati relation to have up to 40\% violators and not be considered as failing for the data set.  Therefore, the scatters are already being accounted for, and it is overkill to use such a generous limit to perform the test.  If the test is done in this manner, no longer can allowances be made for any violators (or, more precisely, there needs to be less than 0.3\% violators).  Even by the groups own tests, there are violators on the order of a few percent, depending on the test.  This is an unacceptable violator rate considering they are violating a limit from the three-sigma deviation from the model.  Finally, another question that arises is that the bursts we see all seem to be biased in one direction.  If we were seeing the result of measurement scatter about the Amati relation, we should expect to see an equal fraction of bursts well above the limit line.  Instead, we see that for almost all data sets, the bursts are systematically in one direction from the limit.

The Amati relation will certainly see improvement in these tests in the future.  With increasing number of \textit{Swift} bursts with spectroscopic redshifts, it will undoubtedly eventually lie right in the middle of the \textit{Swift} data set.  Even then, the Amati relation will be failing for our best data sample, the BATSE data.  Our argument is that there are undeniable systematic effects at play that are causing the Amati relation, and therefore even these `improvements' would be fairly meaningless as we would still see systematic differences in where bursts are observed in the diagrams.  Therefore, the Amati relation is simply not good for making any kind of predictions, cosmological or otherwise.

Perhaps the simplest disproof of the Amati relation is simply that the violator fraction is greatly too high in most data sets.  And perhaps the simplest proof that the Amati relation is caused by selection effects is the large differences between the various $S_{bolo} - E_{peak,obs}$ diagrams for the many detectors.

\section{Conclusions}

The $S_{bolo} - E_{peak,obs}$ diagram has two limit lines, where bursts cannot be below that line if the Amati or Ghirlanda relation holds.  Actually, with the fairly large total uncertainties, substantially larger than the simple measurement errors quoted in the literature, we can expect nearly half of the bursts to be scattered below the Amati limit line.  So a simple test of the Amati relation is whether the {\it average} burst falls below the Amati limit.  (This is similar to the original test proposed by Nakar \& Piran, except that agreement with the Amati relation corresponds to about 40\% violators.)  We apply this test to many burst samples.  The samples of early bursts with spectroscopic redshifts (as originally used to calibrate the Amati relation) pass our test, as does the sample of HETE bursts (even though the scatter about the Amati relation is unusably large).  All other satellites have a large fraction of violators far below the Amati limit line.  This is true whether we look at bursts with or without measured spectroscopic redshifts.  This constitutes a proof that the Amati relation could possibly apply, at best, to only a small and unrecognizable fraction of GRBs.  Indeed, the wide variations in distribution from detector to detector constitute a proof that selection effects must dominate the Amati relation.

We find four selection effects restrict the distribution on all sides.  The best known detector selection effect is the trigger threshold, which produces a roughly horizontal and fuzzy cutoff.  A more subtle and more restrictive selection effect is that for an $E_{peak,obs}$ value to be reported, the burst must be brighter than some threshold, with this threshold rising fast with increasing $E_{peak,obs}$.  These two detector selection effects will cut out bursts that are some combination of faint and hard, with these effects changing greatly from detector to detector.  The third and fourth selection effects operate to restrict the burst population as it appears in the sky.  The third selection effect is that bursts have a log-normal distribution of $E_{peak,obs}$ with the mean value shifting to lower values for faint bursts.  This effect will also reduce the number of detectable bursts that are faint and hard.  The fourth selection effect is that bright bursts are much rarer than faint bursts, as quantified by the usual power-law $\log [N(>P)]-\log[P]$ curve.  The combination of the third and fourth effects means that the bright and soft bursts are  doubly-rare, so that the upper-left side of the $S_{bolo} - E_{peak,obs}$ diagram will be empty.

For a detector with a range of spectral sensitivity and a low detection threshold, the distribution in the $S_{bolo} - E_{peak,obs}$ diagram will extend relatively low, with a large fraction of violators below the Amati limit (like for BATSE).  For a detector with a low energy range of sensitivity and a low detection threshold, the cutoff will be a diagonal line just below the Amati limit.  When combined with the paucity of bright-soft bursts in the GRB population (i.e., those above the Amati limit line), we have a combined selection effect that picks out bursts near the Amati limit.  Such a burst sample would then appear to follow the Amati relation.  Thus, the very strong selection effects for the early bursts with spectroscopic redshifts will create the Amati relation without any need for a physical connection between the $E_{peak,obs}$ and $S_{bolo}$.  That is, the Amati relation is not real, but its appearance in some data sets is simply a result of various selection effects by the detectors and within the GRB population.

With these strong results, the Amati relation should clearly not be used for purposes of cosmology, as has been previously done by many groups.  In particular, BATSE provides the strongest evidence for not using the Amati relation for cosmological purposes.  We note that our groups have never used the Amati relation for any cosmological purpose (e.g., Schaefer 2007; Xiao \& Schaefer 2011).

Our community has expected a connection between $E_{peak}$ and energy for some time (e.g. Lloyd et al. 2000).  With the Amati relation clearly unsuitable, the other possible interpretation is non-physical correlation between $E_{peak}$ and $E_{iso}$.  As demonstrated above, this relation is extremely weak (e.g., see Table 1).  That is, for any given detector, some $\langle\log\frac{E^{2.04}_{peak}}{S_{bolo}}\rangle$ of observed bursts can be used as a sort of intercept for such a correlation with the given dispersion.  For some detectors, this empirical correlation can serve as a useful description of the data, and indeed this is what gave rise to the proposed Amati relation.  However, for many detectors, this interpretation is not useful, however, because the given dispersions are as wide as the data itself, so any such statements regarding a correlation between the two would be largely meaningless.

We emphasize that the failure of the Amati relation in no way carries any implications for any other GRB luminosity relation.  The fault of the Amati relation can be viewed as if it is merely a version of the Ghirlanda relation except that the beaming correction is unknown, so isotropic emission was assumed. The result, however, is that the Amati relation is biasing itself towards some average of whatever the beaming factors of the calibrating bursts are.  All the other GRB luminosity relations do {\it not} involve beaming corrections, and the known physics of the beaming is already accounted for in the physics derivations of these laws.  The Ghirlanda relation is in essence just a conservation of energy statement, while the other luminosity relations (all involving the peak flux, not the fluence) just involve relativistic effects in the visible region of colliding jets.  Indeed, most of the other GRB luminosity relations were {\it predicted} from the physics and then later confirmed.  In all, the failure of the Amati relation is zero evidence for the validity of the other relations (many of which were confirmed predictions) and there are good physical reasons to know that they are valid physical laws for GRBs.

{}

\begin{deluxetable}{ccccccc}
\tablewidth{0pc}
\tabletypesize{\small}
\tablecaption{Demographics of the Data Samples}
\tablehead{\colhead{Data Set} &  \colhead{\#$_{w/\textit{z}}$} & \colhead{$\xi_{w/\textit{z}}$\tablenotemark{a}} & \colhead{$\langle\log\frac{E^{2.04}_{peak}}{S_{bolo}}\rangle_{w/\textit{z}}$\tablenotemark{b}} &  \colhead{\#$_{w/o \; \textit{z}}$} & \colhead{$\xi_{w/o \; \textit{z}}$\tablenotemark{a}} & \colhead{$\langle\log\frac{E^{2.04}_{peak}}{S_{bolo}}\rangle_{w/o \; \textit{z}}$\tablenotemark{b}}}
\startdata
Ideal, no scatter & \ldots & 0\% & 8.95$\pm$0.29 & \ldots & 0\% & 8.95$\pm$0.29 \\
Ideal, with scatter & \ldots & $\sim$40\% & 9.33$\pm$0.61 & \ldots & $\sim$40\% & 9.33$\pm$0.61 \\
Amati et al.  2006 & 50 & 34\% & 8.90$\pm$0.56 & 0 & \ldots & \ldots \\
Schaefer 2007 & 27 & 41\% & 8.95$\pm$0.57 & 0 & \ldots & \ldots \\
BATSE & 0 & \ldots & \ldots  & 1654 & 93\% & 10.18$\pm$0.88  \\
HETE  & 12 & 33\% & 8.67$\pm$0.62 & 24 & 54\% & 9.05$\pm$0.84 \\
\textit{Swift} &  25 & 76\% & 9.42$\pm$0.47  & 46 & 85\% & 9.46$\pm$0.45 \\
Suzaku &  7 & 100\% & 9.77$\pm$0.74 & 25 & 92\% & 10.28$\pm$0.87 \\
\textit{Swift}-Suzaku & 28 & 86\% & 10.01$\pm$1.01 & 38 & 74\% & 9.63$\pm$0.85 \\
Konus &  33 & 73\% & 9.42$\pm$0.58 & 64 & 78\% & 9.68$\pm$0.87 \\
\textit{Beppo}-SAX & 10 & 90\% & 9.36$\pm$0.39 & 119 & 90\% & 9.51$\pm$0.39 \\
\enddata
\tablenotetext{a}{$\xi$ is the fraction of bursts that violate the 'Amati limit' of $1.13\times10^9 \unit{keV}^{2} \unit{erg}^{-1} \unit{cm}^2$ for the Amati energy ratio $E^{2}_{peak} / S_{bolo}$.}
\tablenotetext{b}{For the Amati relation, the quantity $E^{2.04}_{peak} / S_{bolo}$ should never exceed $1.1\times10^9 \unit{keV}^{2.04} \unit{erg}^{-1} \unit{cm}^2$, so even with normal observational scatter a sample of GRBs should not have $\langle\log\frac{E^{2.04}_{peak}}{S_{bolo}}\rangle>9.04$ in appropriate units.  With a reasonable distribution of burst distances, the limit will be even smaller.  So this column provides a measure of the disagreement with the Amati limit for a sample of bursts.  The RMS scatter for $\log\frac{E^{2.04}_{peak}}{S_{bolo}}$ is given after the average value, so this can give a measure of the scatter of the distribution.}
\label{MasterTable}
\end{deluxetable}

\clearpage

\clearpage

\clearpage

\clearpage

\clearpage

\clearpage

\clearpage

\clearpage

\clearpage

\clearpage

\clearpage

\clearpage

\clearpage

\clearpage

\end{document}